\documentclass[pdftex]{aa}

\usepackage{natbib}
\usepackage{graphicx}
\usepackage{txfonts}
\usepackage{amssymb}
\usepackage{amsmath}
\usepackage{upgreek}
\usepackage{setspace}
\usepackage{ragged2e}

\usepackage{color}
\definecolor{referee}{rgb}{0.0,0.0,0.0}
\definecolor{referee2}{rgb}{0.0,0.0,0.0}

\begin{document}

\title{The secondary eclipses of WASP-19b as seen by the ASTEP\,400 telescope from Antarctica}

\author{L. Abe\inst{1}, I. Gon\c{c}alves\inst{1}, A. Agabi\inst{1}, A. Alapini\inst{2}, T. Guillot\inst{1,3}, D. M\'{e}karnia\inst{1}, J.-P. Rivet\inst{1}, F.-X. Schmider\inst{1}, N. Crouzet\inst{4}, J. Fortney\inst{3}, F. Pont\inst{2}, M. Barbieri\inst{1}, J.-B. Daban\inst{1}, Y. Fante\"\i-Caujolle\inst{1}, C. Gouvret\inst{1}, Y. Bresson\inst{1}, A. Roussel\inst{1}, S. Bonhomme\inst{1}, A. Robini\inst{1}, M. Dugu\'{e}\inst{1}, E. Bondoux\inst{1}, S. P\'{e}ron\inst{1}, P.-Y. Petit\inst{1}, J. Szul\'{a}gyi\inst{1}, T. Fruth\inst{5}, A. Erikson\inst{5}, H. Rauer\inst{5}, F. Fressin\inst{6}, F. Valbousquet\inst{7}, P.-E. Blanc\inst{8}, A. Le van Suu\inst{8}, S. Aigrain\inst{9}}

\offprints{L. Abe,\\
      e-mail: Lyu.Abe@unice.fr}

\institute{
    Universit\'{e} de Nice Sophia-Antipolis, Observatoire de la C\^ote d'Azur, CNRS UMR 7293, 06108 Nice Cedex 2, France
\and
    School of Physics, University of Exeter, Stocker Road, Exeter EX4 4QL, United-Kingdom
\and
    Department of Astronomy \& Astrophysics, University of California, Santa Cruz, CA 95064 USA
\and
    Space Telescope Science Institute, Baltimore, MD 21218, USA
\and
    DLR Institute for Planetary Research, 12489 Berlin, Germany
\and
    Harvard-Smithsonian Center for Astrophysics, Cambridge, MA 02138, USA
\and
    Optique et Vision, 6 bis avenue de l'Est\'erel, BP 69, 06162 Juan-Les-Pins, France
\and
    Observatoire de Haute-Provence, Universit\'e d'Aix-Marseille \& CNRS, 04870 Saint Michel l'Observatoire, France
\and
    Department of Physics, University of Oxford, Oxford OX1 3RH, United Kingdom
}

\date{DRAFT: \today }


\abstract
{}
{
The ASTEP (Antarctica Search for Transiting ExoPlanets) program was originally
aimed at probing the quality of the Dome C, Antarctica for the discovery and
characterization of exoplanets by photometry. In the first year of operation of
the 40\,cm ASTEP\,400 telescope (austral winter 2010), we targeted the known
transiting planet WASP-19b in order to try to detect its secondary transits in
the visible. This is made possible by the excellent sub-millimagnitude
precision of the binned data. }
{
The WASP-19 system was observed during 24 nights in May 2010. Once brought
back from Antarctica, the data were processed using various methods, the best
results being obtained with an implementation of the Optimal Image Subtraction
(OIS) algorithm. }
{
The photometric variability level due to starspots is about $1.8\%$ (peak-to-peak), in line with the SuperWASP data from 2007 ($1.4\%$) and larger than in 2008 ($0.07\%$).  We find a rotation period of WASP-19 of $10.7\pm 0.5$\,days, in agreement with the SuperWASP determination of $10.5\pm 0.2$ days. Theoretical models show that this can only be explained if tidal dissipation in the star is weak, i.e. the tidal dissipation factor $Q'_\star>3\times 10^7$. Separately, we find evidence for a secondary eclipse of depth \textcolor{referee}{$390 \pm 190$\,ppm} with \textcolor{referee}{a 2.0$\sigma$ significance}, a phase consistent with a circular orbit and a $3\%$ false positive probability. Given the wavelength range of the observations (420 to 950\,nm), the secondary transit depth translates into a day side brightness temperature of \textcolor{referee}{$2690_{-220}^{+150}$}\,K, in line with measurements in the $z'$ and K bands. \textcolor{referee}{The day side emission observed in the visible could be due either to thermal emission of an extremely hot day side with very little redistribution of heat to the night side, or to direct reflection of stellar light with a maximum geometrical albedo $A_{\rm g}=0.27\pm 0.13$.} \textcolor{referee2}{We also report a low-frequency oscillation well in phase at the planet orbital period, but with a lower-limit amplitude that could not be attributed to the planet phase alone, and possibly contaminated with residual lightcurve trends.}
}
{
This first \textcolor{referee}{evidence} of a secondary eclipse in the visible
from the ground demonstrates the high potential of Dome C, Antarctica for
continuous photometric observations of stars with exoplanets. These continuous
observations are required to understand star-planet interactions and the
dynamical properties of exoplanetary atmospheres.
}

\keywords{Instrumentation: photometers -- stars: individual: WASP-19 --
(stars:) planetary systems -- (stars:) starspots -- Planets and satellites:
atmospheres}

\authorrunning{The ASTEP Team}
\titlerunning{ASTEP\,400 Observations of WASP-19}
\maketitle

\defcitealias{Hebb2010}{\textcolor{referee}{H10}}

\section{Introduction}

The observation of primary and secondary planetary transits at multiple wavelengths is a powerful tool to characterize the atmospheres of exoplanets, understand their energy budget, atmospheric dynamics and thermal evolution. The first detections of a secondary transit (\citealt{Deming2005}), of phase variations of a non-transiting planet (\citealt{Harrington2006}) and of a transiting planet (\citealt{Knutson2007}), confirmed theoretical predictions of a strong day-night temperature contrast and powerful zonal winds in close-in exoplanets (\citealt{ShowmanGuillot2002}). However, in spite of a very rapid increase of the number of exoplanets for which these detections have become possible, our understanding of these atmospheres remains sketchy at most. This is in part because most of the detections have been done at infrared wavelengths and have therefore concerned the thermal part of the planetary spectrum, rather than also the visible part which is essential to quantify the amount of irradiation absorbed by these atmospheres. The situation has been changing thanks to observations by the space telescopes MOST (e.g. \citealt{Rowe2008}), CoRoT (e.g. \citealt{Alonso2009a,Alonso2009b}, \citealt{Snellen2009,Snellen2010}) and Kepler (e.g. \citealt{Borucki2009}, \citealt{KippingBakos2011}), but progress has been slow due to their
intrinsic limitations in accessible targets.

We present near-continuous observations of the WASP-19 star-planet system acquired at visible wavelengths from the ASTEP\,400 telescope at the Concordia station, Dome\,C, Antarctica. As described by \citet{Crouzet2010}, the Dome\,C has excellent meteorological conditions and near-continuous observations are possible for approximately three months centered on the Southern winter solstice (June 21st). This makes it an ideal site for long-duration observations such as
those required to characterize the phase curves of exoplanets.

The target, WASP-19, is hosting one of the exoplanets with the shortest orbital period known thus far. It was discovered as part of the WASP survey in the southern hemisphere (\citealt{Hebb2010}, \citetalias{Hebb2010} hereafter). Its visible magnitude of 12.3 and declination of about $-45.66^\circ$ made it a perfect target for observation from Concordia. The atmospheric properties of exoplanet WASP-19b are beginning to be well-characterized in the infrared thanks to observations of its secondary eclipses at 1.62 and 2.09\,$\upmu$m using VLT/HAWK-I (\citealt{Anderson2010}, \citealt{Gibson2010}, respectively), at 3.6, 4.5, 5.8 and 8.0\,$\upmu$m using Spitzer (\citealt{Anderson2011}), and recently in the z' band at $0.9\,\upmu$m with the NTT/ULTRACAM (\citealt{Burton2012}). The interest in the system is further enhanced due to its short orbital period (0.79 days) and Jupiter-like mass (1.17\,M$_{\rm Jup}$) which implies that its survival is critically dependent on the efficiency of tidal dissipation in its solar-type host star (0.97\,M$_\odot$). Any constraint on the orbital properties of the system, in particular orbital eccentricity and spin period of the star can help us to better understand its dynamical evolution, fate, with potential implications for the entire population of close-in exoplanets. The article is organized as follows: We first present the observational setup and processing of the data. The results are presented in Sect.\,\ref{sect:results}. In Sect.\,\ref{sect:analysis}, we discuss how these results compare with other well characterized exoplanetary atmospheres, and how
these measurements can be applied to understand the dynamical evolution of the systems.

\section{Observations and data processing}

The ASTEP (Antarctica Search for Transiting Exo-Planet) program comprises the ASTEP-South 10\,cm refractor (\citealt{Crouzet2010}) and the ASTEP\,400 40\,cm telescope (\citealt{Daban2010}) that was used for the WASP-19 observations reported in the present paper. The goal of the observations was not only to try to detect a signature of the secondary transit, but also the possible reflected light phase variations of the planet, and the stellar luminosity variation reported in \citetalias{Hebb2010}. We therefore observed the target star continuously for 24 days.

\subsection{Observation log}

WASP-19 was one of our first ASTEP\,400 program targets. It was observed between April $30^\mathrm{th}$ and May $23^\mathrm{rd}$ 2010 (UTC). We summarize the basic observation parameters and statistical information in Table\,\ref{Table:WASP19_CampaignStats}. \textcolor{referee}{The information concerning} duty cycle (instrument and data recording) is measured relative to the
overall time span of the WASP-19 campaign.

The temperature ranged from -78$^\circ$C to -44$^\circ$C with an average value of -67$^\circ$C. On the instrument side, we already noticed ice forming on the primary and sometimes on the secondary mirror. At that time we did not have any permanent defrosting system installed, so a daily manual defrosting operation was mandatory. Icing on the optics did of course affect the absolute photometry, but since it affected all PSFs in the field, we could obtain good lightcurves by
careful data processing (see below).

\begin{table}
\caption{\label{Table:WASP19_CampaignStats} The ASTEP\,400 WASP-19 observations parameters and statistics.}
\centering
\begin{spacing}{1.3}
\begin{tabular}{lrl}
    \hline
    \hline
    \textcolor{referee}{Data}               &  \textcolor{referee}{Value} & \textcolor{referee}{Notes}\\
    \hline
    Average temperature                     &          -68$\pm 7^\circ$C  & \\
    Exposure time                           &                     130\,s  & single frame\\
    \textcolor{referee}{Frame periodicity}  &                     150\,s  & median\\
    PSF FWHM                                &         4 arcsec (4.3 pix)  & median\\
    Max. sun elevation                      & -10$^\circ$ to -6$^\circ$   & daily\\
    Instrument Duty cycle$^\spadesuit$      &                       61\%  & \\
    Overall data duty cycle                 &                       58\%  & $\sim$14hr daily\\
    Science only data duty cycle            &                       48\%  & \\
    \hline
\end{tabular}
\end{spacing}
\justifying
\noindent$^\spadesuit$ fraction of time of instrument operation including science and calibration data, telescope and systems overheads.\\
\end{table}

\subsection{ASTEP\,400 transmission}
The ASTEP\,400 telescope consists of a 400\,mm Newton telescope designed to withstand the harsh conditions in Antarctica and a sophisticated field corrector. The beam is reflected by 2 aluminum-coated mirrors, before being sent to the thermalized field corrector where it first crosses a double lens designed to maximize both thermal isolation and image uniformity. The beam is then split by a dichroic so that the blue part is used for guiding and the red part of the light is used for the science camera. On its way to the science camera, the beam passes again through three lenses plus the CCD glass window. The science camera is an FLI Proline with a KAF 16801E, $4096\times4096$ pixels front-illuminated CCD and a 16 bits analog to digital converter (see
\citealt{Daban2010} for a complete description of the instrument).

\begin{figure}
    \centering
    \begin{minipage}[c]{\columnwidth}
        \centering\includegraphics[width=\textwidth]{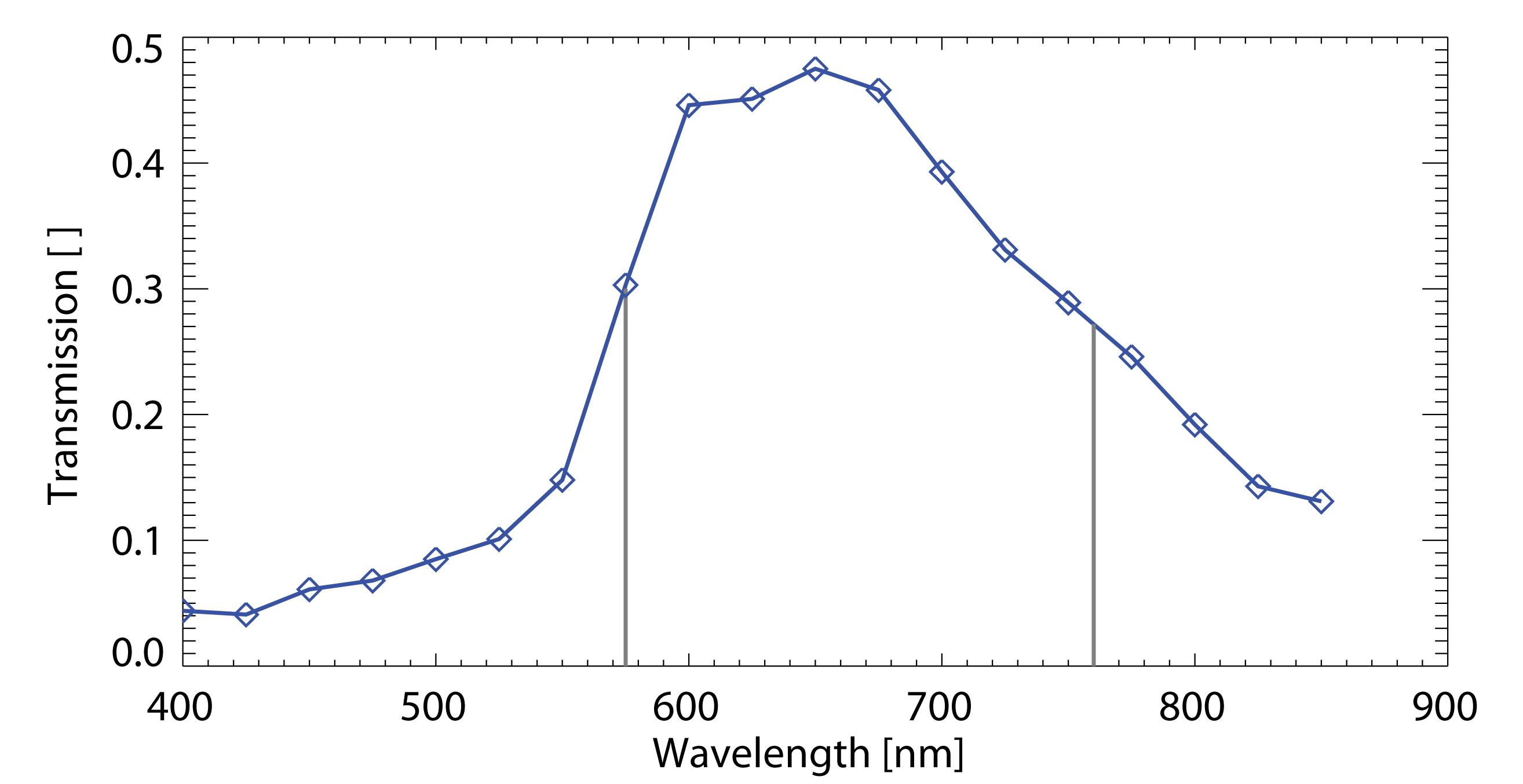}
    \end{minipage}
    \caption{Transmission in the ASTEP\,400 camera as a function of wavelength (in nm). The vertical lines indicate the range of wavelengths within which 65\% of the flux of a $T_{\rm eff}=5500\,$K blackbody is transmitted, i.e. 575 to 760\,nm.}
\label{fig:Instrument.Transmission}
\end{figure}

Figure~\ref{fig:Instrument.Transmission} shows the final transmittance for the science image as a function of wavelength calculated using Zemax and technical data obtained from the constructors for each element and the CCD camera. Most of the signal is transmitted in the red. In order to estimate more precisely our wavelength coverage for science images, we assume that transmission stops below 300\,nm and above 1000\,nm. We then calculate the range of wavelengths that correspond to a 65\% and a 95\% transmission of a $T_{\rm eff}=5500\,K$ blackbody, corresponding to that of WASP-19. The result is 575 to 760\,nm and 420 to 950\,nm, respectively. (In the case of a uniform irradiation,
we find that 65\% of the flux would be transmitted between 580 and 780\,nm.)

\subsection{Data reduction\label{sect:data.reduction}}
Each of the ASTEP\,400 frames is pre-processed using standard bias and dark corrections. Flat field correction turned to be a little more complicated since we had to deal with the sky-concentration effect (see e.g. \citealt{Andersen1995}). \textcolor{referee}{We used satellite trails and star brightness statistics to infer the effect of the sky-concentration, and confirm that the bell-shaped halo at the center of our images was a purely additive component. We removed it by performing a $32\times32$ median smoothing on a $256\times256$ reduced copy of the frame. The enlarged version of this smoothed image is then subtracted} \textcolor{referee2}{from} \textcolor{referee}{the original frame. This process had to be applied to each individual frame since the sky-concentration changed slightly over time. The flat field frames were obtained with the same process.}

The ASTEP\,400 data pipeline exists under two different implementations: a custom IDL code using classical aperture photometry routines from the well known IDL astronomical library, and the Miller-Buie implementation of the \textit{Optimal Image Subtraction} algorithm (OIS, \citealt{Miller2008}). The first implementation is used for processing the data on-site at the Concordia station, whilst the second one was developed later.

\textcolor{referee}{The OIS core algorithm (see Appendix\,\ref{Appendix.A}) is similar} \textcolor{referee2}{to the ISIS algorithm (\citealt{Alard1998})} \textcolor{referee}{with the difference that in ISIS the convolution kernels are made of combinations of 2-D Gaussian functions, whilst in the OIS, these kernel functions are completely arbitrary (the \textit{optimal} ones with respect to the subtraction residual).} \textcolor{referee2}{The superiority of ISIS-type algorithms over aperture photometry has been clearly demonstrated for example in \citet{Montalto2007}.} The OIS requires much more computational time, but also significantly improves the scatter in \textcolor{referee2}{all} our lightcurves compared to aperture photometry \textcolor{referee}{by a factor $\sim$1.5 in RMS} \textcolor{referee2}{see Appendix\,\ref{Appendix.Compare.OIS.AP}}\textcolor{referee}{. There is also an improvement of the noise structures (red noise) that makes the OIS data globally cleaner}. \textcolor{referee2}{Specifically for the WASP-19 target which is amongst the brightest stars in that field, the gain is 20\%.} Quite surprisingly very few papers (e.g. \citealt{Gillon2007}) report on the systematic photometric biases that is intrinsic to the image subtraction methods. A description of this effect and a way to compensate is discussed in Appendix\,\ref{Appendix.A}. \textcolor{referee2}{In the case of WASP-19, the RMS gain compared to aperture photometry is about 20\% (see folded lightcurve comparison in Appendix\,\ref{Appendix.Compare.OIS.AP}).}

\begin{figure*}
    \centering
    \begin{minipage}[c]{\textwidth}
        \centering\includegraphics[width=\columnwidth]{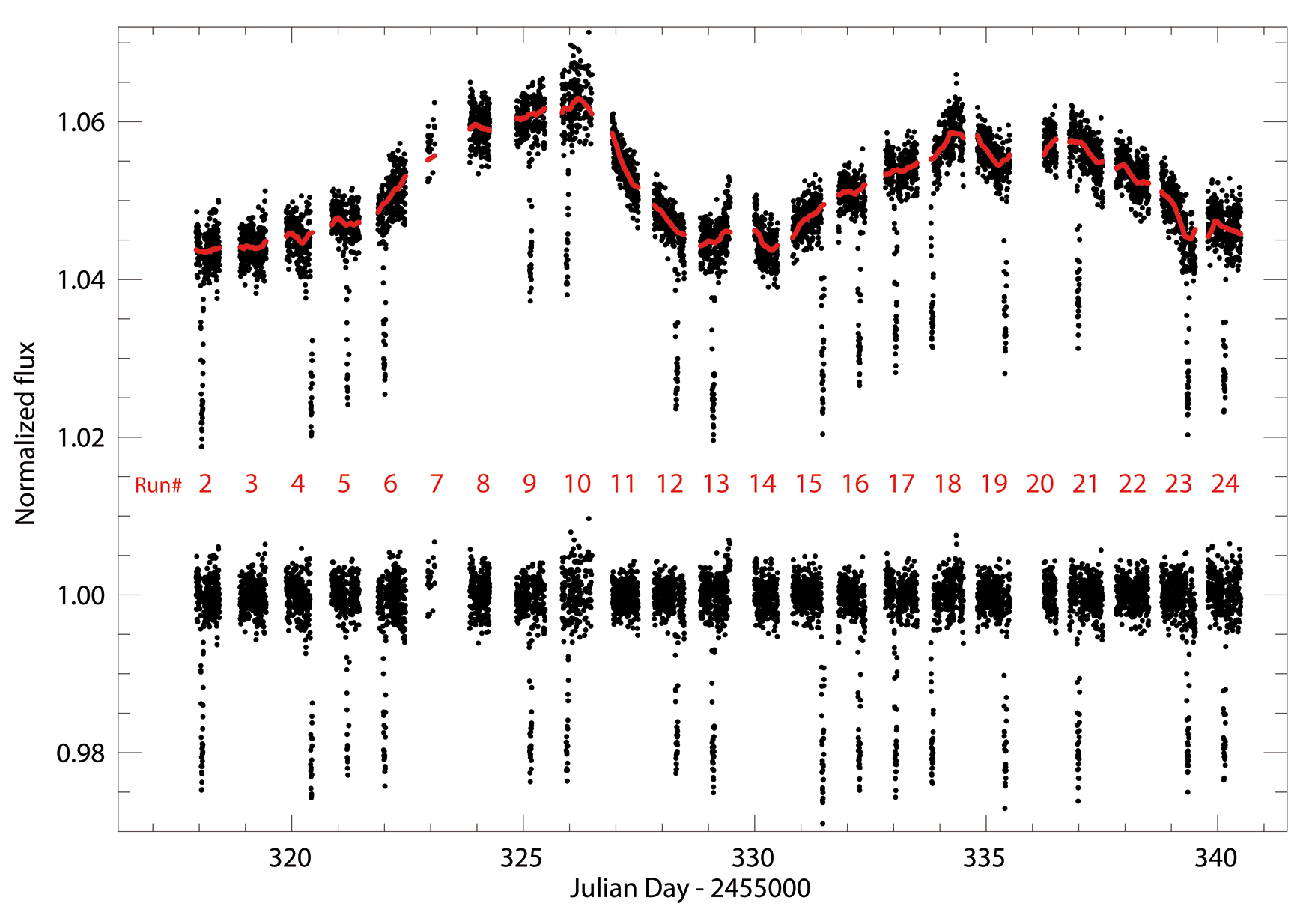}
    \end{minipage}
   \caption[example]
    { \label{fig:WASP19_SeasonalLC} (Top) 23-days lightcurve of the WASP-19 star showing its
variability (shifted for clarity). \textcolor{referee}{The stellar variability is over-plotted in
red, including a final filtering step to remove residual systematics (sliding median of 6.25 hours
in width).} (Bottom) Same as the top lightcurve, but with the stellar variability removed (see text
for details).}
\end{figure*}

\textcolor{referee}{The OIS is intrinsically removing all global variations for all stars. The remaining systematics are calibrated using the following methodology: about 3000 stars are pre-selected as not being noticeably varying (using an unsharp mask filter on the raw lightcurves and keeping stars with less RMS scatter). Amongst these 3000 remaining stars, only 10 stars are kept as references for each of the analyzed star. The selection criteria consists in keeping those 10 stars that minimize the most the RMS scatter after the lightcurve is normalized. We have compared our calibration procedure with other methods such as SysRem \citep{Tamuz2005}, or the method described in \citep{Mazeh2009} without noticeable differences.}

Prior to applying primary transit and secondary eclipse model fitting (see Sect.\,\ref{sect:primary.parameters} and \ref{sect:secondary_eclipse}), the obtained lightcurve (Fig.\,\ref{fig:WASP19_SeasonalLC}--top) needed to be filtered to remove the stellar variations. A calibration lightcurve (excluding primary transit windows) was obtained by a temporal median filter, that was then smoothed by applying a Fourier-space filter. This smoothed lightcurve was subtracted \textcolor{referee}{from} the original one. The obtained residual lightcurve (flattened) was median-filtered again. The temporal width of this last filter was obtained by running a simulation where the primary transit was simply replaced by artificial 500\,ppm-deep transit events. The optimal filter width was found to be 6.25\,hr, as it was selected to maximize the number of artificial eclipse detections. The final calibration lightcurve is completed by interpolating the out-of-transit smoothed lightcurve at the primary transit locations. The ``flattened'' lightcurve (Fig.\,\ref{fig:WASP19_SeasonalLC}--bottom) is obtained by subtracting the smoothed variability lightcurve to the original data.

\subsection{Data quality}

\textcolor{referee}{For each night, the \textit{pink} noise -- gaussian and correlated noise combined -- was analyzed with a variant of the \citet{Pont2006} prescription, taking into account the preprocessing of the lightcurves. As described in \citep{Gillon2006}, it consists in computing the variance of the pink (\textit{white} + \textit{red}) noise on the lightcurves, and on a smoothed version of them (the bin size corresponding to the duration of the transit). Each noise variance follows the relation $\sigma_\mathrm{pink}^2 = \sigma_\mathrm{white}^2/N + \sigma_\mathrm{red}^2$ where $N$ is the number of points used for smoothing the data. The solution of this system of two equations are the white and red standard deviations $\sigma_\mathrm{white}$ and $\sigma_\mathrm{white}$. The different nights were found to vary in stability, with the hour-timescale pink noise at best $430\pm65$\,ppm, median 700\,ppm and worst $955\pm250$\,ppm, comparable to very good ground-based transit photometry. The uncertainties on these values were evaluated by generating artificial data with corresponding white and red noises statistics. The red noise uncertainty is 200\,ppm in average.}

\begin{figure*}
    \centering
    \begin{minipage}[c]{\textwidth}
        \centering\includegraphics[width=\textwidth]{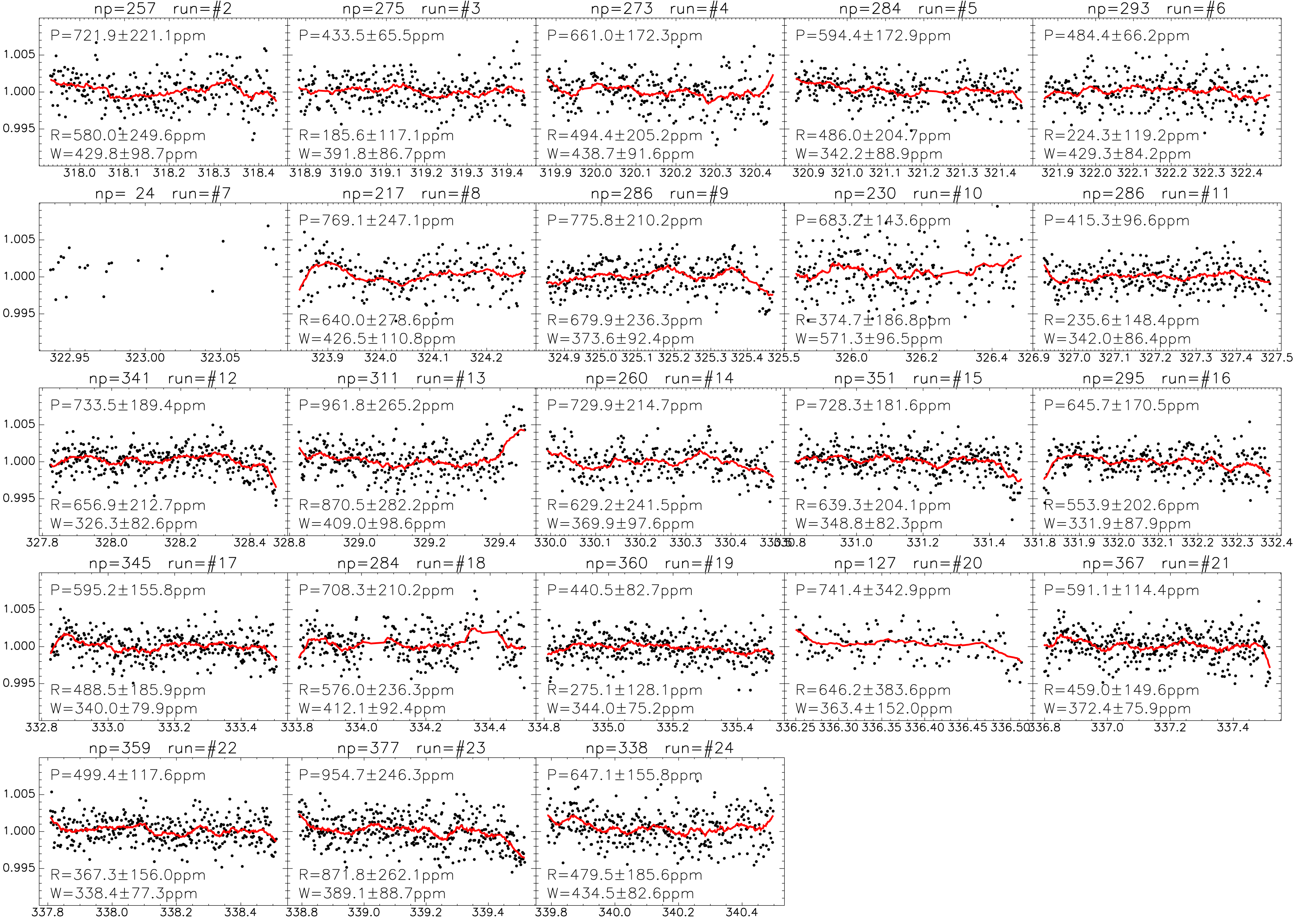}
    \end{minipage}
    \caption{\label{fig:red.pink.white.noise}\textcolor{referee}{Summary of noise statistics for each night computed on the lightcurve residuals (data subtracted from the model) for each complete run (or equivalently for each observing night). Labels `\textsf{P}', `\textsf{R}' and `\textsf{W}' stand for \textit{pink}, \textit{red} and \textit{white}, respectively. The run (or night) number as well as the number of considered data points (\textsl{np}) is indicated above each frame. The abscissa correspond to HJD-2455000.}}
\end{figure*}

\textcolor{referee}{The result of this analysis (Fig.\,\ref{fig:red.pink.white.noise}) shows that 3 nights are ``not-so-good'', with a pink noise over 1 hour around 0.1\,\% (night 7 has too few points to compute relevant statistics, but was kept as it adds information for the stellar variability). Most nights are ``good'', with a pink noise over 1 hour around 0.7\,mmag, and a few ``very good'' nights, with a pink noise over 1 hour around 400\,ppm. These numbers are typical to current ground-based photometry. The best ground-based transit photometry now reaches 0.4\,mmag (KeplerCam, VLT infrared), the ``good'' nights are also around 0.7, and more casual observations are above 1\,mmag.}

During the first night, we noticed that there was a switch of guiding star  \textcolor{referee}{from the good one to another one}. \textcolor{referee}{Taking into account this first night produced odd results (strong trends) when processing for the photometry with the OIS pipeline. In addition, the fraction of observing time when the telescope was tracking the correct guide star was too short to affect our result significantly so we deliberately discarded all data points from that night.} In the rest of this paper, all results are obtained from the 23 remaining nights.

\section{Results\label{sect:results}}

\subsection{Stellar variability\label{sect:stellar_variability}}
\textcolor{referee}{Before the ASTEP\,400 campaign, the only observations with a long-enough duration to extract the stellar spin period were those of \citetalias{Hebb2010}, who found a periodicity of $10.5\pm0.2$ days}. Our data also exhibit a periodic oscillation with an amplitude of about 0.9\% as well as a period of $10.7\pm0.5$ days, compatible with the 2007 data from \citetalias{Hebb2010}. Although our lightcurve is much more precise, we did not record enough periods to improve the result from \citetalias{Hebb2010}.

The lightcurve exhibits a clear asymmetry. This asymmetry can actually also be guessed \emph{by eye} in the \citetalias{Hebb2010} data where the oscillation maximum seems slightly offset from a purely sinusoidal signal (Fig.\,6 of their paper). In order to confirm this, we have processed the publicly available SuperWASP WASP-19 data and performed several tests showing that the asymmetry is present in 2007 and comparable in shape (or trend) to what we obtained in 2010. In 2008, we find a change in the asymmetry, although with an amplitude about twice fainter than in \textcolor{referee2}{2007} and 2010. \textcolor{referee}{This appears to result directly from the presence of stellar spots and their progressive evolution}.

\subsection{Primary transit\label{sect:primary_transit}}

\subsubsection{Transit parameters\label{sect:primary.parameters}}

\begin{figure*}
    \begin{minipage}[c]{\textwidth}
        \centering\includegraphics[width=\textwidth]{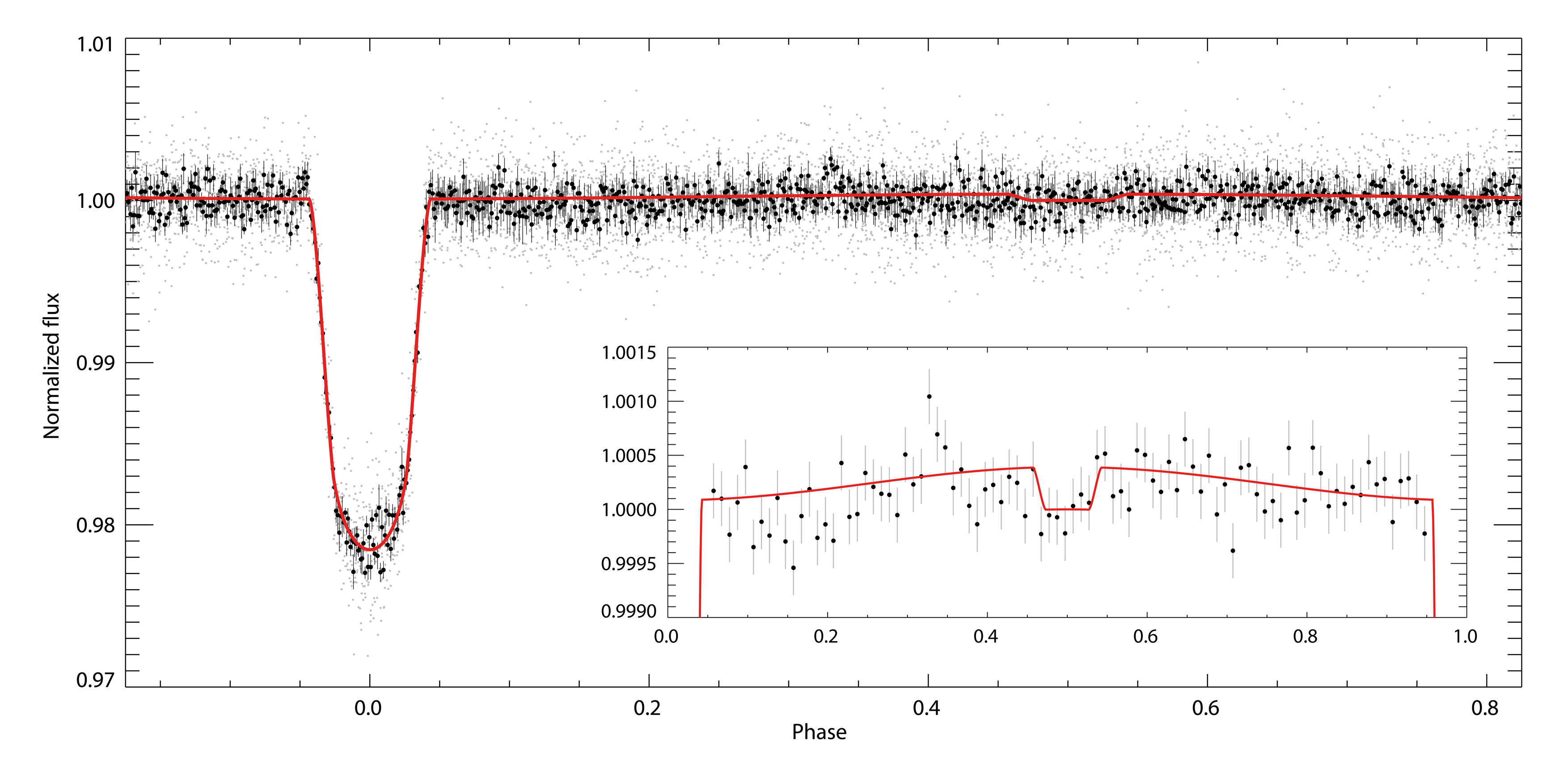}
    \end{minipage}
    \caption{Phase-folded lightcurve of WAPS-19b, where the gray and blue dots correspond to the original data, and to the binned data at 0.001 phase units (blue), respectively. \textcolor{referee}{The black line is the best fit to the transit (see Table~\ref{Table:WASP19_TransitParam} for detailed primary transit parameters). The} \textcolor{referee2}{inset} \textcolor{referee}{frame is the normalized flux scale zoom on the secondary eclipse (full phase coverage in abscissa) with a bin size of 0.01 in phase -- see Sect.\,\ref{sect:secondary_eclipse} for detailed secondary eclipse parameters.}}
    \label{fig:primary_transit}
\end{figure*}

The phase-folded transit on Fig.\,\ref{fig:primary_transit} was fitted using the \citet{Mandel2002} analytical transit formulation and an IDL implementation of the Levenberg-Marquardt algorithm (LMFIT) to find the best fit (black curve on Fig.\,\ref{fig:primary_transit}). We assume a circular orbit to define the timing of the secondary eclipse \textcolor{referee}{in line with the previous secondary eclipse measurements}. The stellar limb darkening was adjusted with a quadratic law. The obtained transit parameters of WASP-19b are summarized in Table \ref{Table:WASP19_TransitParam}, where $P$ is the orbital period, $T_\mathrm{0}$ the reference mid transit time, $R_p$ the radius of the planet, $R_{\star}$ the radius of the star, $a$ the orbit semi-major axis, $i$ the orbital plane inclination, $b$ the impact parameter, $\gamma_1$ and $\gamma_2$ the quadratic limb-darkening coefficients, $e$ the eccentricity, and $\omega$ the argument of periapsis. $\gamma_1$ and $\gamma_2$ were derived from previous studies (\citealt{Hebb2010,Hellier2011}) and tabulated data from \citet{Claret2011}. However, the dependency of the fit on these parameters is very low at this level of photometric precision.

\begin{table}
\caption{\label{Table:WASP19_TransitParam} Best fit transit parameters from the ASTEP\,400 observations.}
\centering
\begin{spacing}{1.3}
\begin{tabular}{lccc}
    Parameter            & Value          & Error           \\
    \hline
    \hline
    $P$ [d]              &      0.788840  & \tiny{(fixed)}  \\
    $T_\mathrm{0}$ [HJD] & \textcolor{referee}{2455317.26998}  & \textcolor{referee}{$\pm$0.00006}    \\
    $R_p/R_{\star}$      &         \textcolor{referee}{0.141}  & \textcolor{referee}{$\pm$0.001}      \\
    $a/R_{\star} $       &         \textcolor{referee}{3.560}  & \textcolor{referee}{$\pm$0.036}      \\
    $i$ [$^\circ$]       &          \textcolor{referee}{79.6}  & \textcolor{referee}{$\pm$0.3}        \\
    $b$                  &         \textcolor{referee}{0.642}  & \textcolor{referee}{$\pm$0.023}      \\
    $\gamma_1$           &          0.42  & \tiny{(fixed)}  \\
    $\gamma_2$           &          0.29  & \tiny{(fixed)}  \\
    $e$                  &             0  & \tiny{(fixed)}  \\
    $\omega$             &             0  & \tiny{(fixed)}  \\
    \hline
\end{tabular}
\end{spacing}
\end{table}

\subsubsection{Fluctuations in transit depth and timing}

The dispersions in transit length, timing, and depth, are compatible with the level of pink noise, assuming no intrinsic variations.

\begin{figure}
    \centering
    \begin{minipage}[c]{0.325\columnwidth}
        \centering\includegraphics[width=\columnwidth]{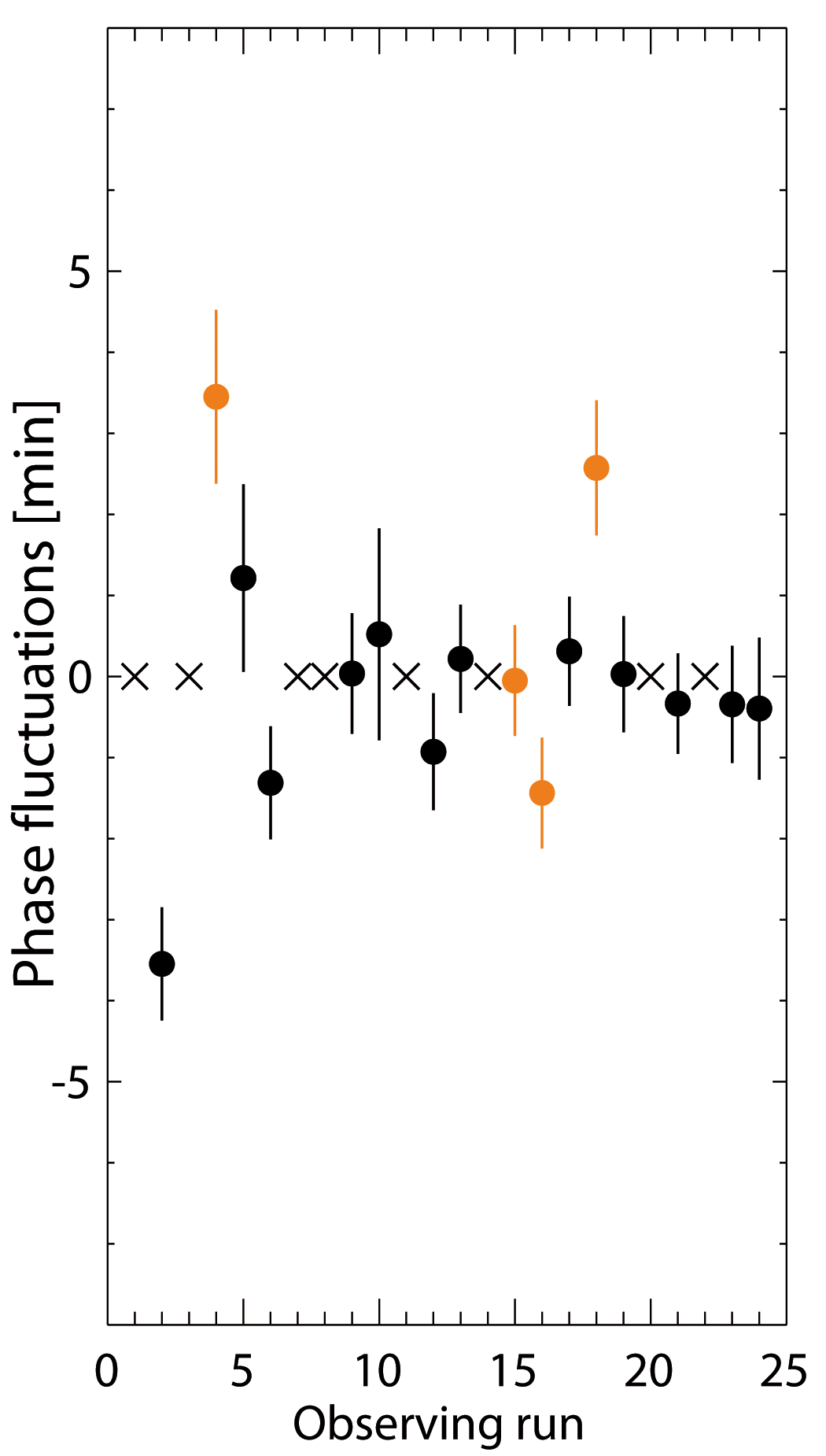}
    \end{minipage}
    \begin{minipage}[c]{0.325\columnwidth}
        \centering\includegraphics[width=\columnwidth]{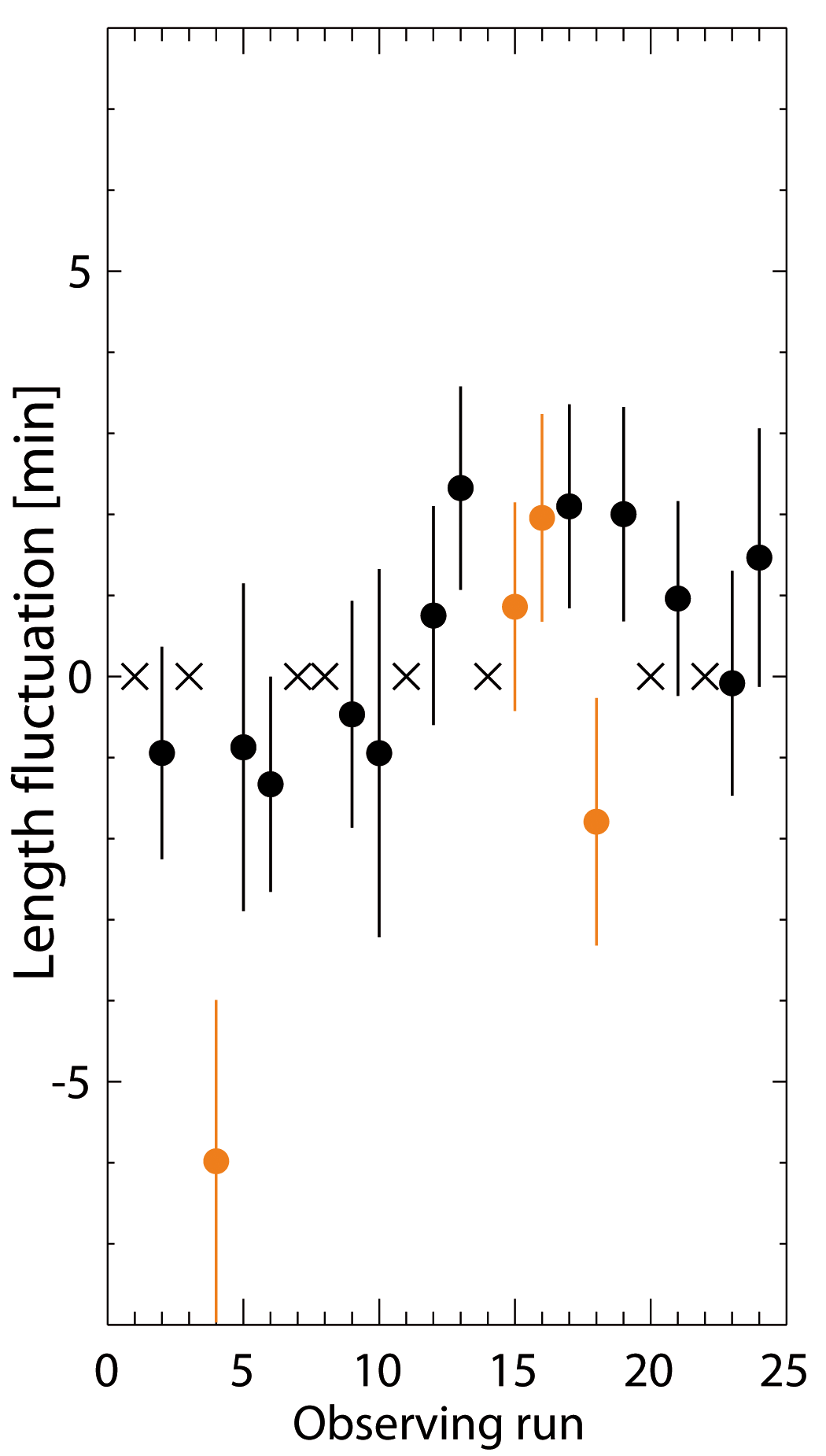}
    \end{minipage}
    \begin{minipage}[c]{0.325\columnwidth}
        \centering\includegraphics[width=\columnwidth]{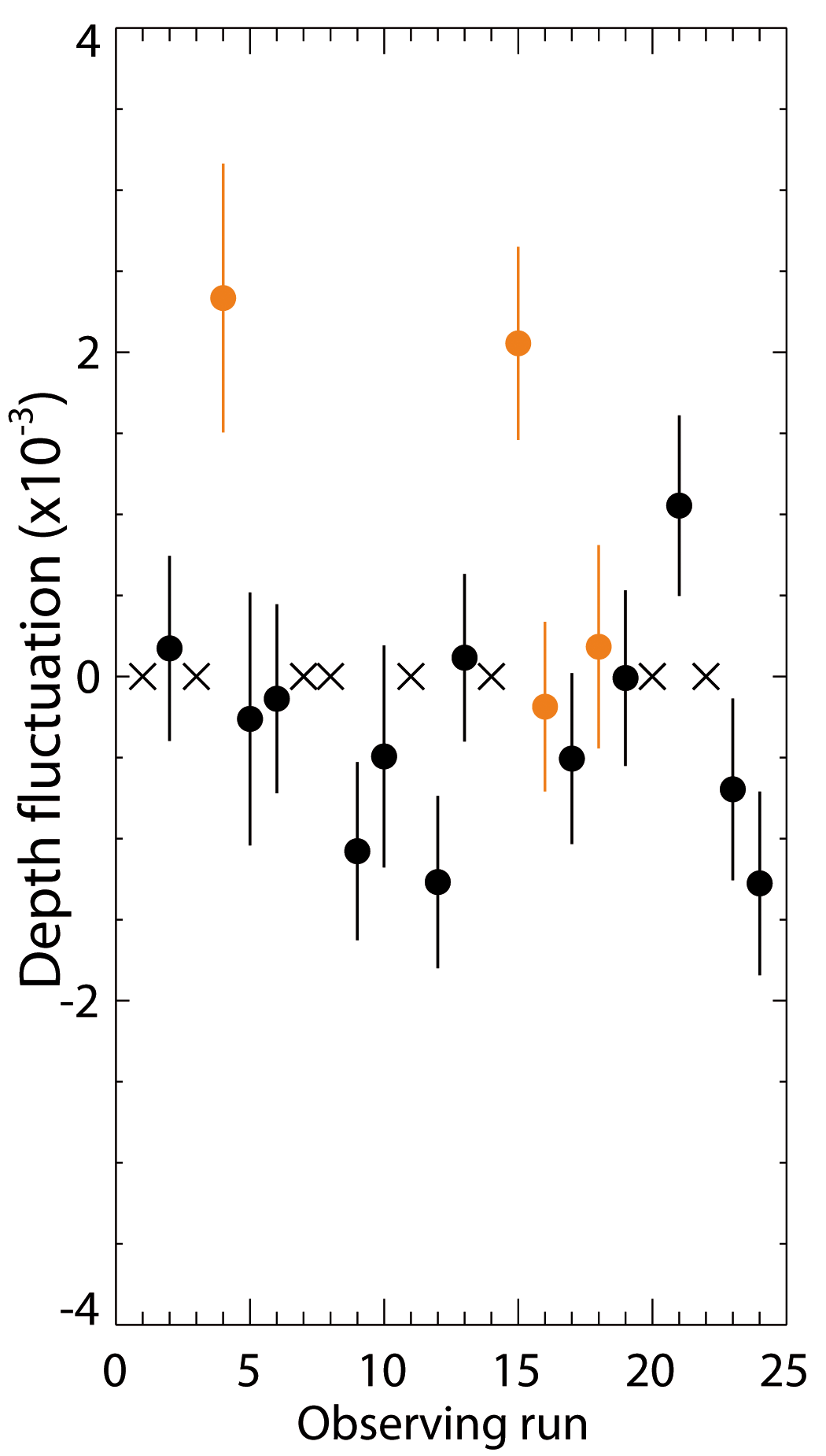}
    \end{minipage}
    \caption{\label{fig:variations}Fluctuations of \textcolor{referee}{the phase timing (left panel),
length duration (middle panel) and depth (right panel)} of individual transits of WASP-19b relative
to the values from the combined transits. The crosses correspond to runs (nights) with no transit
event, and orange point to nights with incomplete primary transit.}
\end{figure}

The primary transits were first fitted altogether to define mean values \textcolor{referee}{(see Fig.\,\ref{fig:day.to.day.transits} in Appendix\,\ref{Appendix.LC})}. We then allowed the depth, timing or duration to vary between the individual transits. Figure\,\ref{fig:variations} shows the resulting dispersion between transits. The uncertainties are calculated from the pink-noise of the respective nights derived above.

We find that the fluctuations in timing and duration are compatible with the error estimates, indicating that our treatment of correlated noise is satisfactory. The point-to-point dispersion of the depth is also compatible with the error estimates.

\subsection{Secondary eclipse and phase effect\label{sect:secondary_eclipse}}

\begin{figure}
    \centering
    \begin{minipage}[c]{\columnwidth}
        \centering\includegraphics[width=\columnwidth]{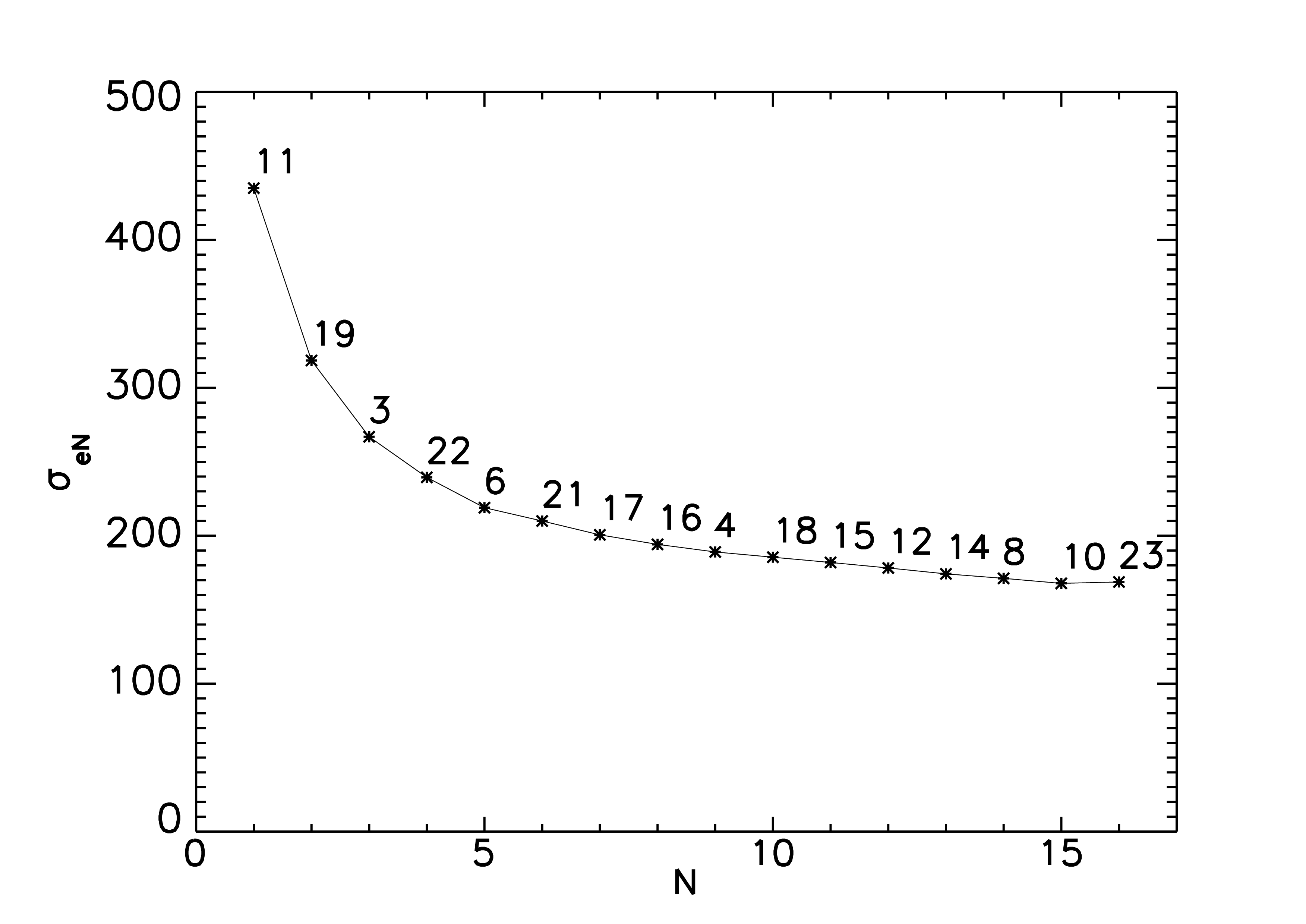}
    \end{minipage}\\
    \begin{minipage}[c]{\columnwidth}
        \centering\includegraphics[width=\columnwidth]{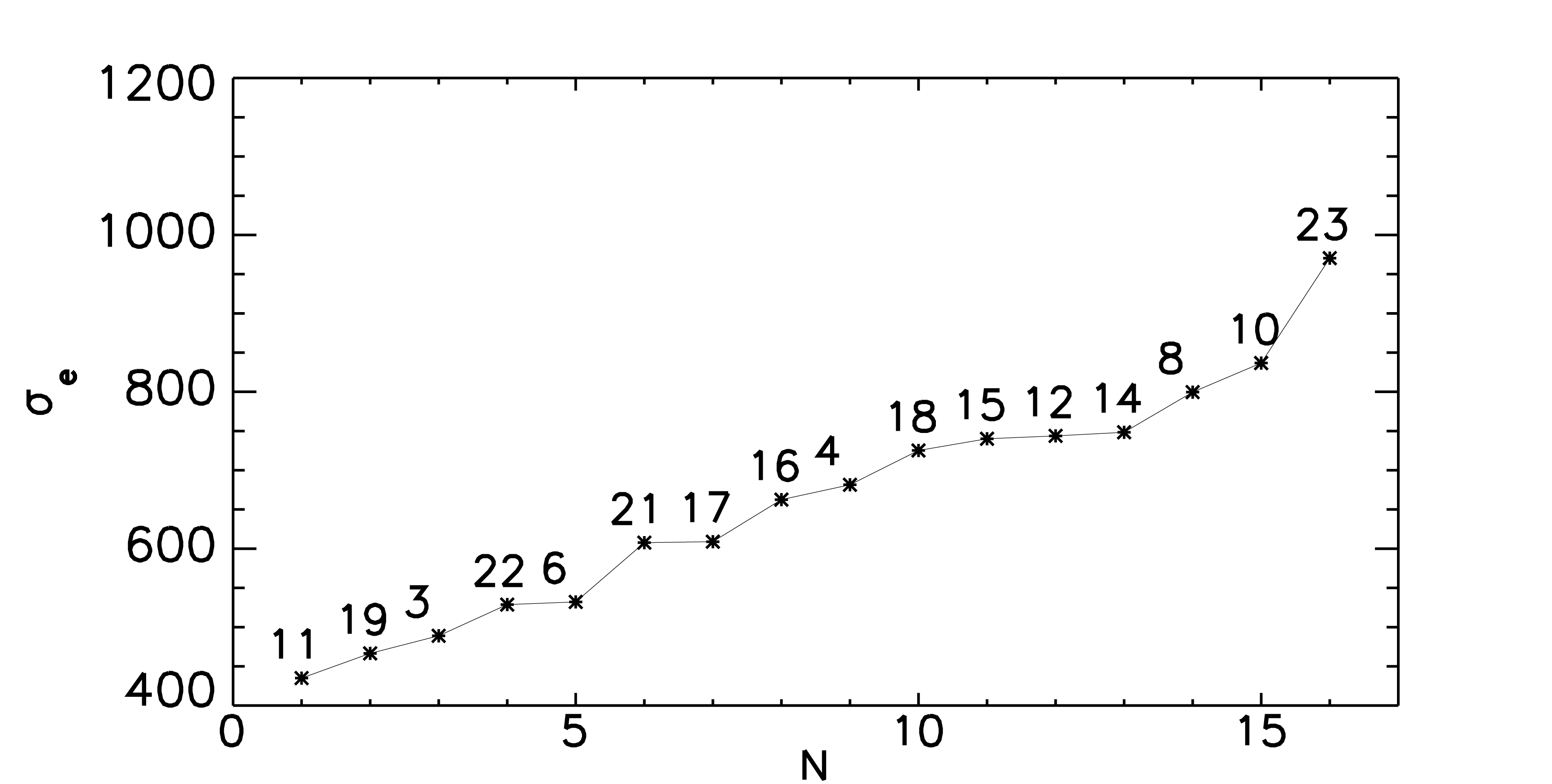}
    \end{minipage}
    \caption{\label{fig:snr.vs.nights}\textcolor{referee}{Standard deviation of the folded lightcurve for 1 hour bins against the number of folded nights (from best to worst). The numbers along the curve correspond to each of the newly added night number}\textcolor{referee2}{: the first point corresponds to a single night whilst the latest corresponds to the 23 nights. The lower panel shows the $\sigma_{e_k}$ value for each night.}}
\end{figure}

\textcolor{referee}{The secondary eclipse is detected on the folded lightcurve, where an approximately synchronous phase effect is also evidenced (inset of Fig.\,\ref{fig:primary_transit}).} \textcolor{referee2}{In order to determine whether to consider all of the observing night, or only specific good ones, we analyzed the behavior of the noise at the eclipse location for an increasing number of $N$ nights. In order to determine the \textit{goodness} of nights, we introduce the eclipse noise $\sigma_{e_k}$,
\begin{equation}
\sigma^2_{e_{\,k}} = \frac{\sigma_{w_{\,k}}^2}{n_{\,k}} + \sigma^2_{r_{\,k}} \nonumber
\end{equation}
\noindent where the $\sigma_{w_{\,k}}$ and $\sigma_{r_{\,k}}$ values are taken from Fig.\,\ref{fig:red.pink.white.noise}. $\sigma_{e_k}$ is estimated for each observing night at the eclipse location, thus taking into account the number $n_k$ of data points at the bottom of the eclipse event. The cumulative noise $\sigma_{e_N}$ in the folded lightcurve is then defined by,}
\begin{equation}
\sigma_{e_{N}} = \sqrt{\frac{1}{n^2} \sum\limits_{k=1}^{N} n^2_k \sigma^2_{e_k}}\nonumber
\end{equation}
\noindent \textcolor{referee2}{with}
\begin{equation}
n = \sum\limits_{k=1}^{N} n_k \,\,.\nonumber
\end{equation}
\noindent \textcolor{referee2}{$\sigma_{e_{N}}$ is plotted on Fig.\,\ref{fig:snr.vs.nights}, and shows that the eclipse noise always improves with increasing number of nights. The plot also displays the cumulative white noise, as well as the value of $\sigma^2_{e_k}$ for each additional night (in a non-cumulative way).}

The lightcurve shown in fig.~\ref{fig:primary_transit}\ was obtained from all the 23 nights, but we checked that the actual measured features did not originate from the few, noisier observing runs only.

After pre-processing the data, we obtain a secondary eclipse fit with a \textcolor{referee}{$390\pm100$\,ppm} depth, using the same Levenberg-Marquardt algorithm as in Sect.\,\ref{sect:primary.parameters}. Note that the $\pm100$\,ppm error only corresponds to the model fitting uncertainty, and should not be confused with the uncertainty on the real depth of the eclipse (detailed below). Considering previous works at other wavelengths (\citet{Gibson2010}) we performed the fit of the secondary with a fixed phase of 0.5 consistent with a circular orbit.

\begin{figure}
    \centering
    \begin{minipage}[c]{\columnwidth}
        \centering\includegraphics[width=\columnwidth]{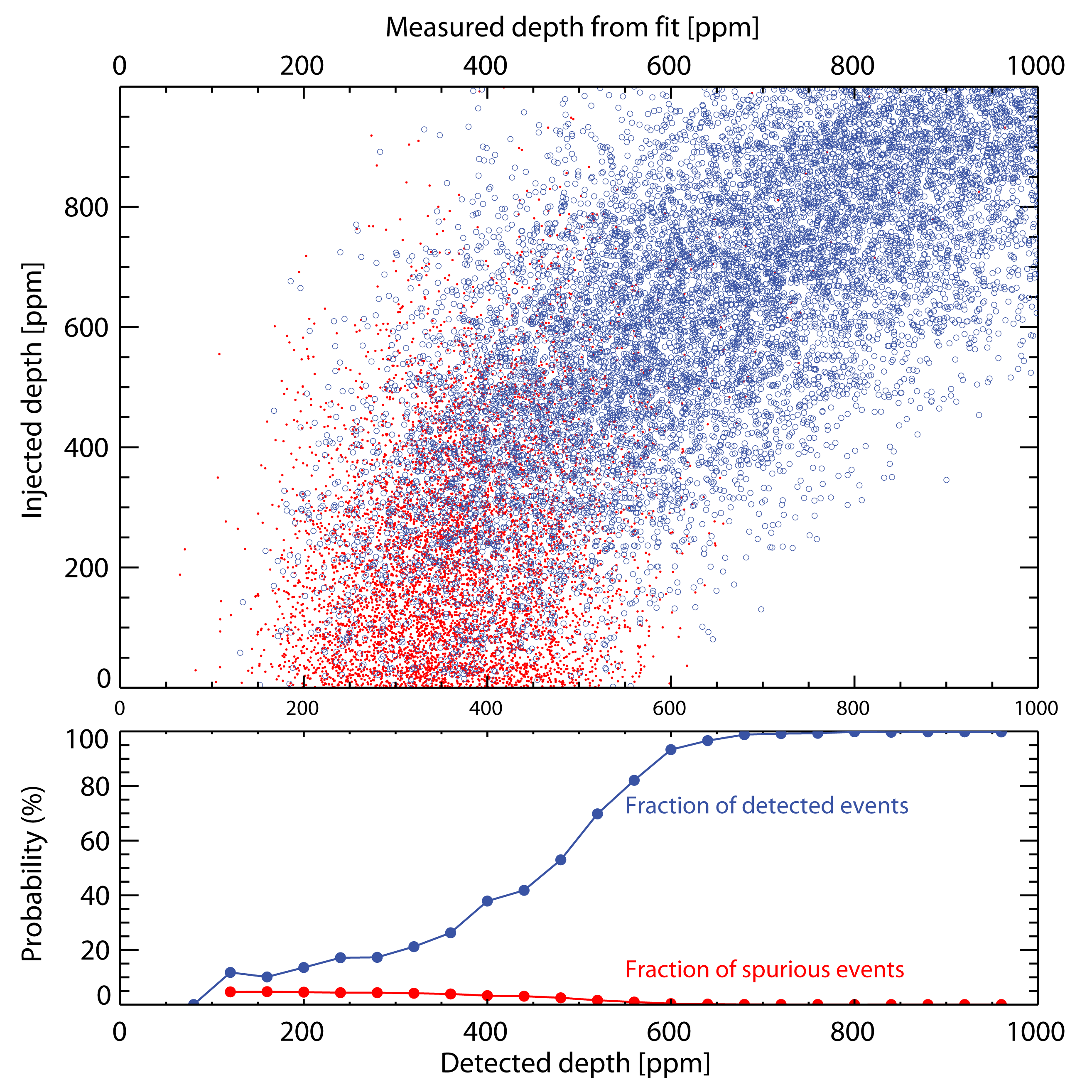}
    \end{minipage}
   \caption[example]
    { \label{fig:detection.stats} (top) Result of the Monte-Carlo simulation points (see text for details). The blue circles show the detected events (i.e. with the phase of the fit matching the expected eclipse location $\pm2.5\%$ in phase unit), whilst the red dots show non detected events (i.e. the result of the fit not matching the expected phase location $\pm2.5\%$). (bottom) The detection (blue) and false alarm (red) probability curves derived from the Monte-Carlo simulation. The detection probability -- for a given depth -- is the ratio between the number of detected events (blue circles) over the total number of events (red plus blue).}
\end{figure}

In order to probe the reality of this detection, we performed several tests. Based on the measured points themselves, we carried out a Monte-Carlo simulation consisting in adding artificial secondary eclipse events to the actual data. These events have a random depths as well as random periods (excluding a range containing the primary transit period). Random periods rather than random phase with the WASP-19b period was used in order to scramble the fluctuations of the noise (especially the red component) as well as systematics. For each of these new data sets, an eclipse detection fit is performed so that we can determine the number of detected events, i.e. the detected events which phase matches the introduced one, without constraints on the depth (what we are looking for). The tolerance for the detected phase is $\pm 2.5\%$, implying a false event detection probability of 5\% assuming that these random false events are uniformly distributed in phase. This tolerance ($\pm2.5\%$) was arbitrarily set as the 2.5-$\sigma$ dispersion of the fitted phases at an eclipse depth of 500\,ppm. The results from this simulation are shown on Fig.\,\ref{fig:detection.stats} where 20\,000 realizations have been drawn. This allows us to assess:
\begin{itemize}
\item[\textbullet]{what we call the \textit{detection sensitivity} to known, injected events of a given depth, which is computed as the ratio of the number of detected events over the total number of injected events at this given depth (i.e. a statistic out of an horizontal cut of the top panel of Fig.\,\ref{fig:detection.stats}),}
\item[\textbullet]{and the \textcolor{referee}{ratio of detected (or spurious) events for a given measured depth}, which is computed as the ratio of the number of detected events over the total number of measured events at this depth (i.e. a statistic out of a vertical cut of the top panel of Fig.\,\ref{fig:detection.stats}).}
\end{itemize}

\begin{figure}
    \centering
    \begin{minipage}[c]{\columnwidth}
        \centering\includegraphics[width=\columnwidth]{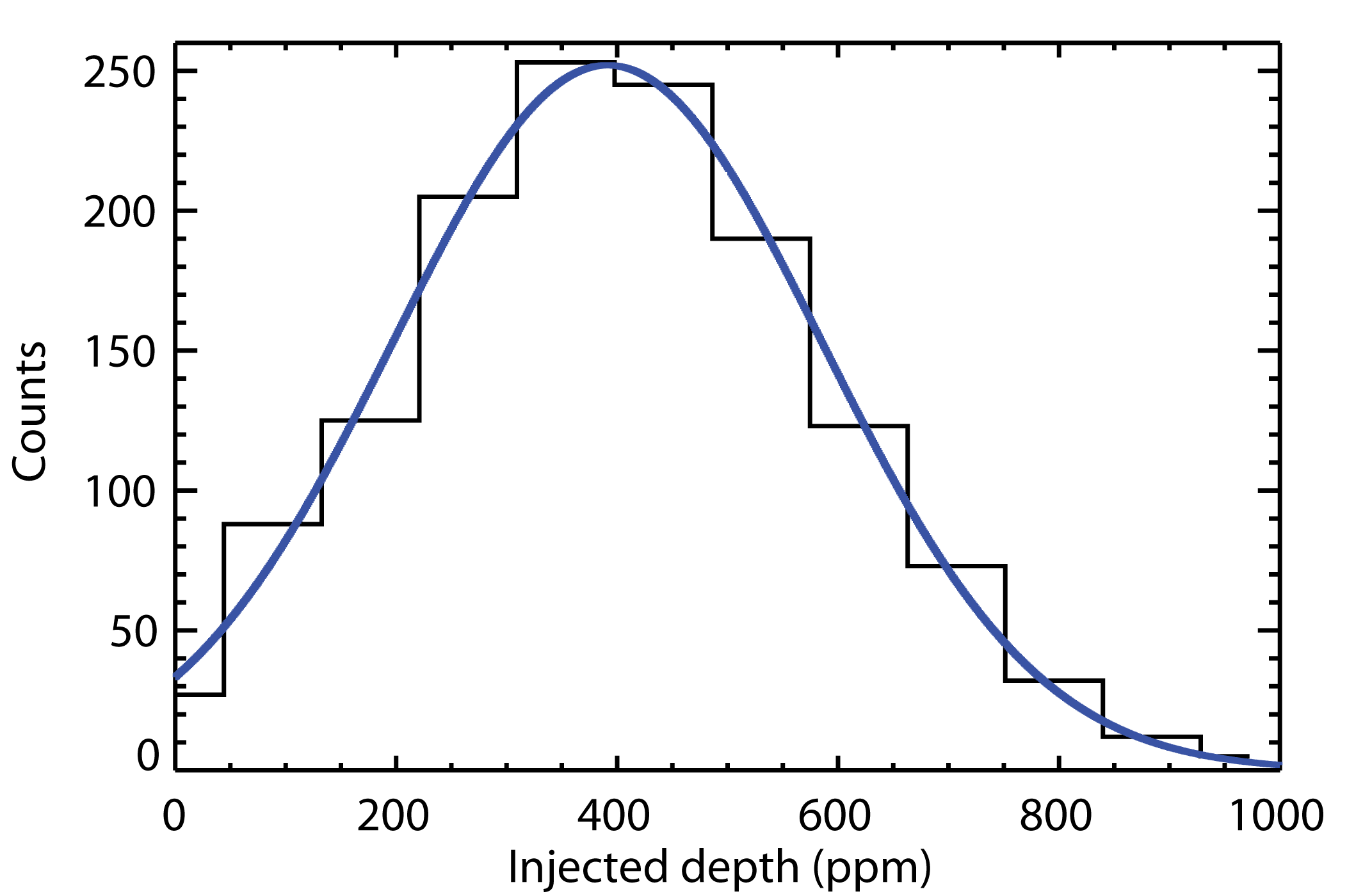}
    \end{minipage}
   \caption[example]
    { \label{fig:detection.historgram} Histogram showing the depth of injected events that were detected in the MC simulation, at the specific measured depth of \textcolor{referee2}{390 (a vertical $\pm40$\,ppm bin through the diagram of Fig.\,\ref{fig:detection.stats}--top)}. The blue curve shows the best-fit Gaussian of the histogram.}
\end{figure}

\noindent For injected eclipses with a depth 390\,ppm, the detection sensitivity at this depth is 69\% (with a bin width of 40\,ppm), while the probability of detecting an eclipse event at this depth is 43\%. However this latter number does not take into account the probability that these events are due to false events.

The uncertainty on the real depth of this event is simply determined by building an histogram of the injected transit depths at the measured depth of 390\,ppm. The result is shown on Fig.\,\ref{fig:detection.historgram}. In order to quantify the average depth and error, we choose to fit a Gaussian to the obtained histogram. The best fit curve gives a depth of \textcolor{referee}{$391\pm5$\,ppm} and a \textcolor{referee}{$\sigma=194\pm4$\,ppm}, which we account for the depth and true uncertainty of the eclipse. We adopt the rounded values of \textcolor{referee}{$390\pm190$\,ppm} as the final result for this work. For the phase of the secondary eclipse, a similar process was applied and lead to a shift in phase (relative to 0.5) found by the model fitting of $45\pm15$\,s, while the error estimate on the ``real'' phase offset is around $\pm30$ minutes. At this level of precision, our result is matching previous works (\citealt{Gibson2010,Hellier2011,Anderson2011}) that find no significant phase shift, or one well below our estimation precision.

\textcolor{referee}{The inset in fig.\,\ref{fig:primary_transit} exhibits a clear  phase-effect that is fitted with a $308\pm90$\,ppm peak-to-valley amplitude for a sine function, and with a phase offset of $-0.04\pm0.03$ relative to the $0.5$ phase. The lightcurve was obtained by filtering the original one to get rid of the stellar variability as described in Sect\,\ref{sect:data.reduction}. The fact is that the last filtering step would introduce an attenuation of the phase effect at this period (if any) by a factor that we estimated to $3.2\pm 1.2$ using a method similar to that of the secondary eclipse Monte-Carlo test.} \textcolor{referee2}{In order to assess the possible origin of this phased function, and understand whether it is real or caused by trends and/or the filtering process, we performed two tests: trying random combination of 10 nights to generate the folded lightcurve, and worked on synthetic data. In the first case, the phase function appeared almost in all cases, so that we are confident that the effect is not caused by specific nights. In the second test, we injected random trends and phased signals to a synthetic lightcurve (same time sampling as the original data) so as to mimic the measured folded lightcurve. These tests suggest that this phased oscillation is compatible with a mix between residual from strongly attenuated trends and an attenuated phased signal that appears to be real (be it of astrophysical origin or caused by systematics), that cannot be clearly disentangled. However, we find that if there is a phased signal, its amplitude would have to be larger ($>500$\,ppm) than the planet phase effect alone to cause such a well-in-phase result.} In conclusion, and in the absence of other work trying to measure the phase effect, we can only be very cautious about its reality in our data, but the correct phasing and symmetry of this hypothetical phase effect is somehow a bit disconcerting. Our observation look promising regarding the possibilities to assess such kind of effect in upcoming observing campaigns. \textcolor{referee}{Note that we also fitted the lightcurve letting the phase effect and eclipse phases as free parameters, as a test of robustness of our secondary eclipse and phase detection. The best fit gives a secondary eclipse depth and phase of 374\,ppm and -0.007, and a phase effect amplitude and phase of 300\,ppm and +0.05.}

\section{Analysis\label{sect:analysis}}

\subsection{Brightness temperatures and albedo}

The analysis of the spectra of exoplanets is complicated by their unknown chemical composition, cloud structure, their inhomogeneous temperature field and complex dynamics. The detection of a secondary eclipse is a measurement of the emission of the day side of the planet, including both thermal emission and reflected light. Disentangling the two generally requires an extremely accurate full phase lightcurve at visible wavelengths, so that both the brightness temperature of the day side (during secondary transit) and the temperature of the night side (during primary transit) can be measured, as exemplified for TrES-2b and HAT-P-7b in the Kepler field (\citealt{KippingBakos2011,Christiansen2010}). Unfortunately, the ASTEP phase lightcurve of WASP-19b does not reach the accuracy required to break the degeneracy. The analysis must therefore rely solely on estimates of the day side brightness temperatures at various wavelengths and comparison with other cases.

The day side brightness temperature $T_{\rm bright}$ of the planet is calculated from the depth $d$ of the secondary transit following \citet{Cowan2011}: \begin{equation} d=\frac {R_{\rm p}^2 \pi B(T_{\rm bright})} {R_{\star}^2 F(T_{\rm eff,\star})}, \label{eq:d} \end{equation} where $R_{\rm p}$ and $R_{\star}$ are the planetary and stellar radius, respectively, $\pi B(T_{\rm bright})$ corresponds to the assumed blackbody emission flux of the planet's day side, and $F(T_{\rm eff,\star})$ is the stellar flux. In this work, we approximate the latter by a theoretical stellar spectrum of the same effective temperature and metallicity from \cite{Hauschildt1998}.

The secondary eclipse depths measured at different wavelengths and the corresponding brightness temperatures \textcolor{referee}{calculated with eq.~(\ref{eq:d}) and for $T_{\rm eff}=5500\,$K, $R_*=0.93\rm\,R_\odot$, $R_{\rm p}=1.3\,R_\mathrm{jup}$ (H10)} are presented in Table~\ref{tab:tbright}. Clearly, the measurements at short wavelengths correspond to smaller depths and are thus more challenging. The corresponding brightness temperatures are always very high and indicate that the day side of WASP-19b is very hot. For comparison, the planet's equilibrium temperature for an assumed 0 Bond albedo (assuming the incoming stellar flux is redistributed uniformly on the planet's surface -- see \citealt{Saumon1996}) is $T_{\rm eq,0}\equiv T_{\rm eff,\star}(D/4R_{\star})^{1/4}= 2000\pm 40\,K$, where $D$ is the star to planet distance.
The maximum effective temperature that would be emitted by the day side if it was a blackbody with no redistribution of heat from the day side to the night side is $T_{\rm max irr,0}\equiv (8/3)^{1/4} T_{\rm eq,0}=2556\pm 51\,$K. Brightness temperatures in excess of that value are possible at selected wavelengths because of opacity variations or in the visible because of reflected light. In the present case, it is interesting to note that the measured temperatures are slightly in excess but still compatible with that maximum effective temperature in bands that span a wide wavelength range, from 0.42 to 2.2\,$\upmu$m.

\begin{table}
\caption{\label{tab:tbright}Measured secondary eclipse depths and brightness temperatures from eq.~(\ref{eq:d}) as a function of wavelength for WASP-19b.}
\begin{center}
\begin{spacing}{1.5}
\begin{tabular}{lrlc}
\hline\hline
Wvl. range ($\upmu$m)   & Depth (ppm)       & $T_{\rm{bright}}$ (K)  & \textcolor{referee}{References}\\
\hline
0.42--0.95              & $390\pm 190$       & \textcolor{referee}{$2690_{-220}^{+150}$}  & \textcolor{referee}{1}\\
0.81--1.01              & $880\pm 190$       & $2640_{-100}^{+90}$   & \textcolor{referee}{2}\\
1.525--1.775            & $2760\pm 440$      & $2740_{-140}^{+130}$  & \textcolor{referee}{3}\\
1.965--2.215            & $3660\pm 670$      & $2680_{-180}^{+180}$  & \textcolor{referee}{4, 5}\\
3.225--3.975            & $4830 \pm 250$     & $2330_{-58}^{+57}$    & \textcolor{referee}{5}\\
4.0--5.0                & $5720\pm 300$      & $2265_{-64}^{+63}$    & \textcolor{referee}{5}\\
5.1--6.5                & $6500\pm 1100$     & $2244_{-230}^{+230}$  & \textcolor{referee}{5}\\
6.55--9.45              & $7300\pm 1200$     & $2240_{-255}^{+252}$  & \textcolor{referee}{5}\\
\hline
\end{tabular}
\end{spacing}
\tablebib{\textcolor{referee}{(1) This work (see Fig.\,\ref{fig:Instrument.Transmission}); (2)
\cite{Burton2012}; (3) \cite{Anderson2010}; (4) \cite{Gibson2010}; (5) \cite{Anderson2011}.} }
\end{center}
\end{table}

he geometrical albedo that would be necessary to explain the measurements in the visible is $A_{\rm g}=d\,(D/R_{\rm p})^2$, i.e. in the case of WASP-19b in the ASTEP\,400 band, we obtain $A_{\rm g}=0.27\pm0.13$. This is to be regarded as an upper limit on the geometrical albedo at these wavelengths. Because of the similarity of the values of the brightness temperatures in the visible and near-infrared, it seems however more likely that the observed signal is due to thermal emission from a very hot day side, with a very inefficient redistribution of heat from the day side to the night side.

\begin{figure}
    \centering
    \begin{minipage}[c]{0.8\columnwidth}
        \centering\includegraphics[width=\columnwidth]{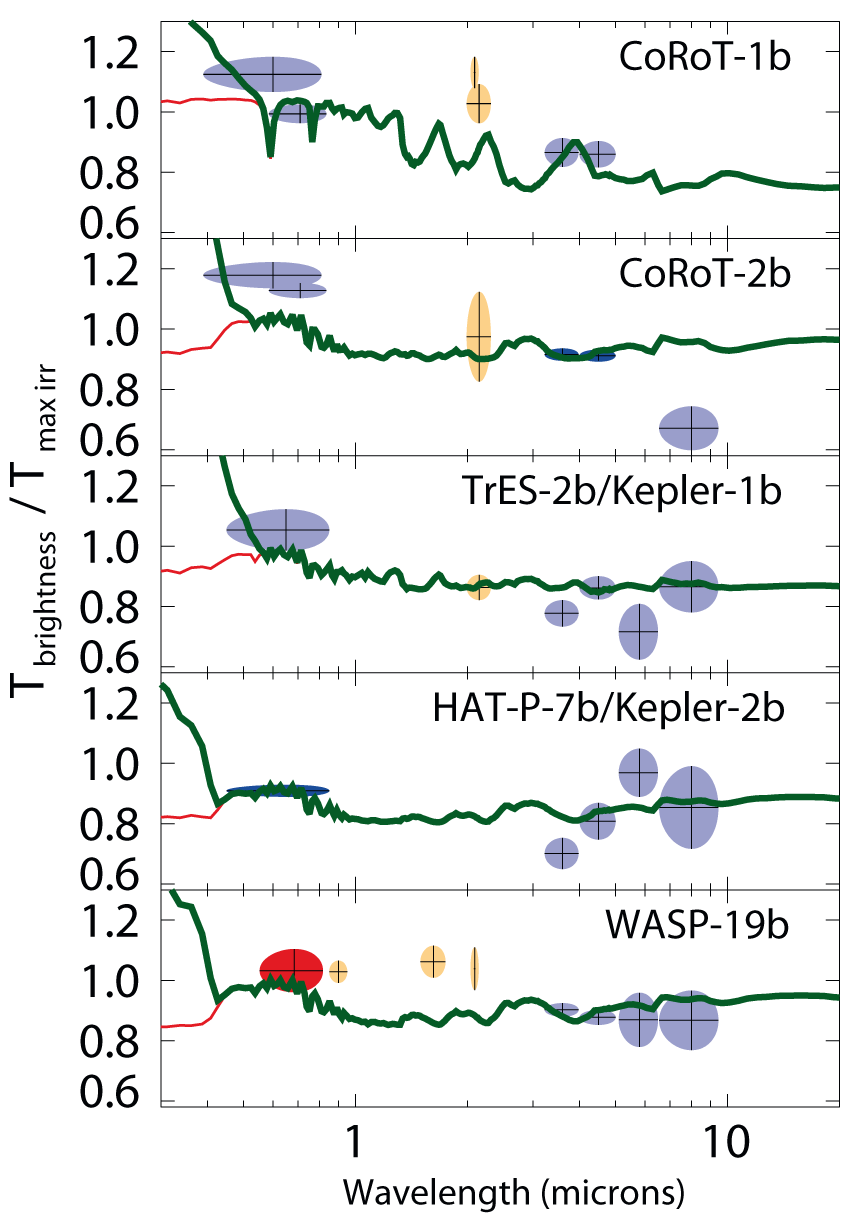}
    \end{minipage}
    \caption{\label{fig:brightness.ratio}Ratio of the brightness temperature to the maximum irradiation temperature (at the substellar point -- see text) versus wavelengths for exoplanets with detected visible and infrared secondary eclipses. Green and red curves correspond to the total modeled spectrum and to the thermal emission only, respectively. The ellipses account both for the bandpasses of the observations (horizontally) and the uncertainties of the measurements (vertically). The colors indicate whether the data originate from space telescopes (blue), mid-latitude ground based facilities (orange), or ASTEP\,400 (red).}
\end{figure}

In order to put the WASP-19 observations in context, we plotted in Fig.\,\ref{fig:brightness.ratio} the exoplanets which show a significant ($>1\sigma$) detection of a secondary eclipse both in the visible and in the infrared, as taken from the compilation of \citet{Cowan2011}. We use the models of \citet{Fortney05,Fortney08} to compare to the observations of these planets.  These are strictly one-dimensional day side models that allow a certainly fraction of absorbed light to be ``lost'' to the night side, which allows the day side $T_{\rm eff}$ to be adjusted.  Here the model fluxes are converted to brightness temperatures.  \textcolor{referee}{Except for CoRoT-1b, all models include TiO and VO as day-side absorbers, which yields day-side temperature inversions. The high K-band fluxes for CoRoT-1b suggest a high temperature in this water opacity window, which argues against a temperature inversion. Solar abundances and equilibrium chemistry are used for all models and no ad-hoc adjustments to chemical mixing ratios were made to yield improved fits}. For the models shown in Fig.\,\ref{fig:brightness.ratio}, the red curves are due to thermal emission, and the green curves are thermal emission plus reflected light.  The green curves turn dramatically upward at short wavelengths due to an increase in the reflected light component, but the geometric albedos at these wavelengths are still quite low.

Clearly, for a number of the planets shown here, the simple models do not capture the true complexity of these atmospheres.  The models generally have redistribution over the day side only, leaving only $\sim$0--20\% of the absorbed energy for the night side.  For WASP-19b, the model has a day side $T_{\rm eff}$=2300 K.  The planet's true day side $T_{\rm eff}$ must however be larger since the model is not bright enough from $\sim$0.8--3\,$\mu$m. In the model we find the optical emission from the planet with ASTEP\,400 is dominated by thermal emission.

\subsection{Tidal evolution}

As shown in \citetalias{Hebb2010}, WASP-19b is at risk of engulfment due to its extreme proximity from its central star. Specifically, the inward migration timescale for a planet of mass $M_{\rm p}$ in a circular orbit of period $P$ around a star of mass $M_\star$, radius $R_\star$ and spin period $P_\star$ is (\citealt{Barker2009}):
\begin{eqnarray*}
\tau_a \approx 12.0 {\ \rm Ma\ } \left(Q_\star'\over 10^6\right) && \left(M_\star\over {\rm M_\odot}\right)^{8/3} \,\,\times\\
&& \left(M_{\rm p}\over {\rm M_{Jup}}\right)^{-1} \left(R_\star\over {\rm R_\odot}\right)^{-5} \left(P\over{\rm1\,day}\right) \left(1-{P\over P_*}\right)^{-1},
\end{eqnarray*}
where $Q_\star'$ is the equivalent tidal dissipation factor in the star (defined as $3/2
Q_\star/k_{2,*}$, where $Q_\star$ is the standard tidal dissipation factor and $k_{2,1}$ its Love number -- see \citealt{Barker2009}). Using the parameters from Table\,\ref{Table:WASP19_TransitParam} we obtain:
\begin{equation}
\tau_a\approx 3.8 {\ \rm Ma\ } \left(Q_\star'\over 10^6\right).
\end{equation}
This implies that either we are extremely lucky to see a system in its few last million years of existence, or more likely the tidal $Q_\star'$ is significantly larger than the assumed $10^6$ typically required to explain the circularization of binary systems (e.g. \citealt{Meibom2005}).

The spin period of the star derived from both \citetalias{Hebb2010} and this work enables to put constraint on the amount of tidal dissipation in the star. As shown for the OGLE-TR-56 system by \citet{Carone2007}, the tides raised by the companion on the central star yield a competition between the orbital migration of the companion, the magnetic braking of the star, and its possible tidal spin-up. As a result, values of $Q_\star$ that are too low may be ruled out because they lead to a spin-up of the star that is too efficient. For the OGLE-TR-56 system, Carone \& P\"{a}tzold find that $Q_\star'>3\times 10^7$.

\begin{figure}
    \centering
    \begin{minipage}[c]{\columnwidth}
        \centering\includegraphics[width=\columnwidth]{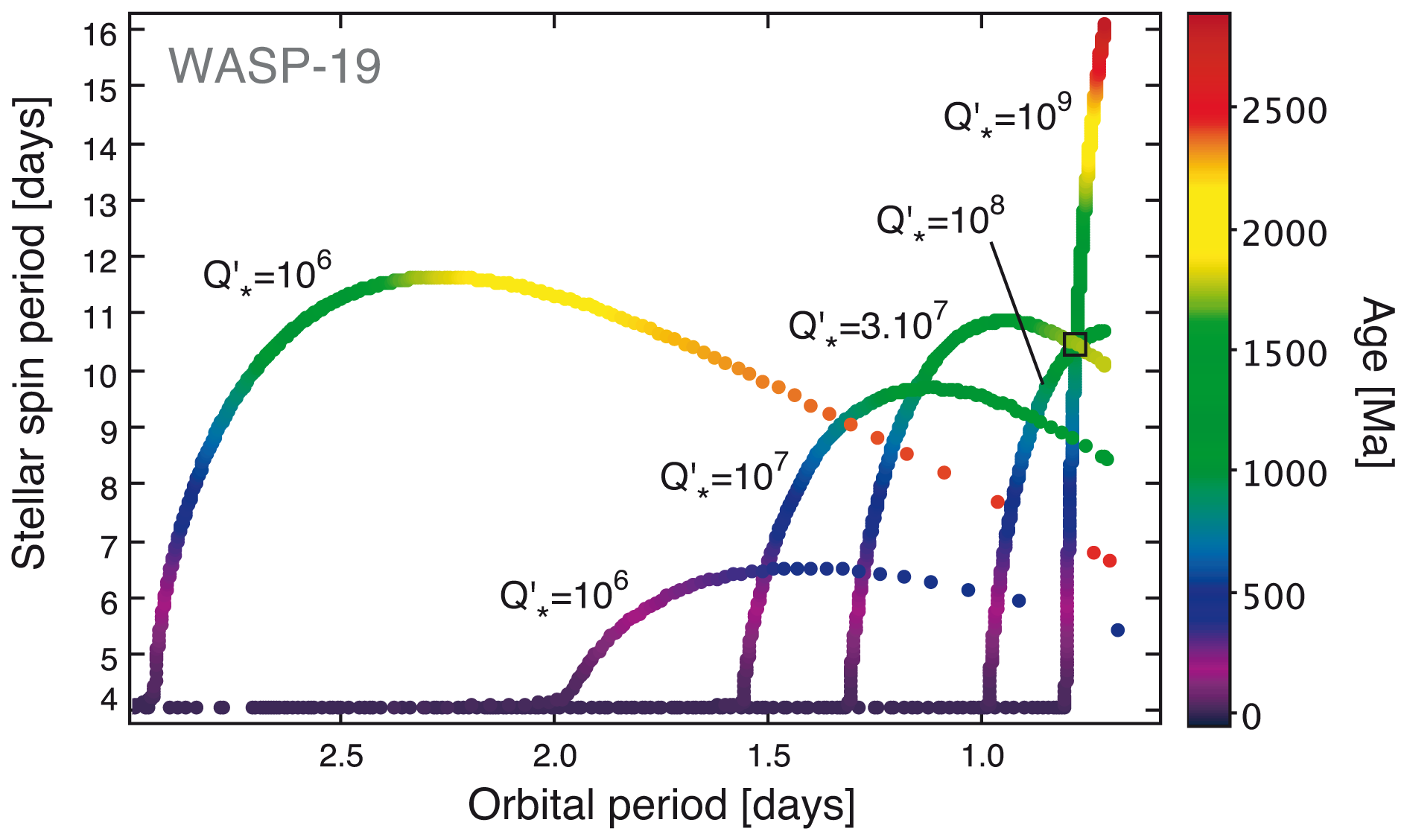}
    \end{minipage}
    \caption{\label{fig:dynamical.evolution}Dynamical evolution of the WASP-19 star-planet system showing the orbital period versus the stellar spin period, in five different cases assuming values of the stellar tidal dissipation factor $Q_\star'$ (see text) ranging from $10^6$ (2 cases) to $10^9$. The corresponding ages, in million years, are shown by the color scale to the right. The observational constraints are represented by a square. Only models with values of  $Q_\star'$ above
$3.10^7$are able to match the observations.}
\end{figure}

We apply the same method to the WASP-19 system, using planar-dynamical evolution equations for the system (\citealt{Barker2009,Guillot2011}), stellar evolution models (\citealt{Morel2008}) and a magnetic braking relation from \citet{Bouvier1997}. We explore the solutions at constant $Q_\star'$ between $10^6$ and $10^9$. We assume an arbitrary initial eccentricity of $0.7$, but even at the highest values of $Q_\star'$ considered, we find that the solutions are independent of this assumption due to the very rapid circularization of the orbit in the first few million years of evolution. The stellar spin is initialized at $4$ days, a value representative of T-Tauri stars (see \citealt{Bouvier1997}). The solutions are found to be weakly dependent of this parameter. For each assumed value of $Q_\star'$, we try to find the value of the initial semi-major axis of the planet that yields a solution in agreement with our observations. The results are presented in Figure\,\ref{fig:dynamical.evolution}. Only values of $Q_\star'$ larger than about $3\times 10^7$ yield both an orbital period and a stellar spin period that match the observations after about $1$ to $1.8$\,Ga of evolution. The latter ages correspond to the lower values of $Q_\star'$ and effectively require a spin-up of the parent star, as indicated by the comparison with other exoplanetary systems and stellar isochrones (see \citealt{Husnoo2012}). These results are in agreement with the values obtained in \citetalias{Hebb2010} with slightly different assumptions. While a more extended exploration of the parameter space would be desirable but beyond the scope of this article, we therefore feel that our results are relatively robust. Surprisingly perhaps, the value obtained is very similar to that obtained for OGLE-TR-56 by Carone \& P\"{a}tzold. While WASP-19 is a G8V type star with a mass slightly smaller than our Sun, OGLE-TR-56 is an F-type star, with an effective temperature around 6100\,K (\citealt{Torres2008}). In both cases, the companions have masses slightly above Jupiter's (1.35\,M$_{\rm Jup}$ for the latter), indicating the dissipation process is weak for Jupiter-mass planets, in line with theoretical predictions of dissipation by internal gravity waves (\citealt{Ogilvie2007,Barker2010}).

\section{Conclusion}

We have presented observations of WASP-19 with ASTEP\,400 obtained during a 24 days observing period through May 2010 that show that extremely stable and precise visible photometry and near-continuous observations are achievable from the Concordia station, Antarctica. The data appear to be of excellent quality, with hour-timescale pink noise averaging $0.7$\,mmag, comparable to very good ground-based transit photometry. The observations confirm that WASP-19 is a relatively spotted star, with $\sim$1\% peak-to-peak photometric variations for this period of observation, and a spin period of $10.5\pm 0.2$\,days.

A detailed analysis of the transit signal yields a detection \textcolor{referee}{of a dip of $390\pm 190$\,ppm at phase 0.5 consistent with the presence of a secondary eclipse. It would correspond} to a brightness temperature of the day side of the planet of \textcolor{referee}{$2690_{-220}^{+150}$}\,K. This is slightly in excess of the maximum possible effective temperature for no redistribution of heat of $2556\pm 51\,$K, but well in line with brightness temperatures obtained in the $z'$ and K bands. This seems to imply that the emission is due to a very bright day side and that heat redistribution is very inefficient. Alternatively, the signal observed in the visible may be due to direct reflection of incoming starlight, in which case we can derive a maximum geometrical albedo $A_{\rm g}=0.27\pm 0.13$. Unfortunately, no atmospheric model appears to correctly account for all observations of the planet so far.

Separately, the ability to fully characterize the system, and in particular the stellar spin period, allows us to put useful constraints on the tidal interactions between the star and its planet and specifically on the tidal dissipation in the star $Q_\star'>~ 3\times 10^7$, a value much larger than that required to explain the circularization of binary stars. The detailed study of close-in giant planets is thus extremely fruitful both to understand planetary atmospheres and tidal interactions between stars and their companions. The ASTEP\,400 detection of a secondary transit in the visible is the first achieved from the ground. \textcolor{referee2}{We also identified a phase function that may originate from the combination of residual trends and a real periodic signal that cannot be clearly disentangled at that point. Further investigations could be carried out with ASTEP\,400, but would require observations for a significantly longer duration.}

\begin{appendix}
\section{OIS photometric correction\label{Appendix.A}}
\subsection{OIS principle}
We recall the OIS principle that consists in subtracting a reference image ($R$) to the current science one ($I$). This reference image is the \textit{best} one in the time series, in the sense of sharpness and sky background level. It is then convolved to an optimal kernel which depends globally on the current science image, and locally to the position in the image (space-varying kernel). Once the reference image is convolved by the optimal kernel ($K$), it has the same average intensity and PSF shapes as the current science image. After subtracting the convolved reference and the science image, we obtain an image with PSF subtraction residuals, whose flux need to be measured by aperture photometry. This operation can be written as,

\begin{equation}
F = F_\mathrm{ref} + \frac{A\,[I - (R \otimes K)]} {\parallel K \parallel}
\end{equation}

\noindent where $F_\mathrm{ref}$ is the flux computed on the reference image with the aperture $A$. With our notation, we could write $F_\mathrm{ref} = A\,R$.

\subsection{Choice of the aperture radius}
The radii of the apertures $A$ (which are different for each star in the reference image) depend on the intensity of the stars and on the global FWHM of all PSFs. \textcolor{referee}{The aperture is allowed to vary for each star so as to minimize the global photometric noise structures in the lightcurves. This is done at the expense of a photometric bias -- as explained below -- that can be compensated using a statistical approach detailed in the next section. The net result is a significant gain in SNR.}

In order to optimize the \textcolor{referee}{photometric} SNR, a simple method consists in selecting an aperture radius by determining all pixels with a level higher than (for example) three times the background noise. By this process, each star in the reference image has its own aperture radius $A$. Note that fainter stars will have smaller reference aperture radii by using this method.

These apertures will give us the reference fluxes $F_\mathrm{ref}$ for each star, but if we use these apertures directly on the subtracted image, they will not take into account the PSF FWHM broadening (it is necessarily a broadening since the reference image is chosen as the sharpest image). A way to account for these FWHM width fluctuations is to convolve the circular reference aperture by the local kernel $K$. The flux is then expressed as,

\begin{equation}
F = F_\mathrm{ref} + \frac{(A \otimes K)\,[I - (R \otimes K)]} {\parallel K \parallel^2}\,\,.
\end{equation}

Apart from broadening the apertures proportionally to the current science FWHM, this aperture convolution has the benefit of `apodizing' the resulting aperture, thus reducing the background noise contribution. To better understand this effect, let us consider the case of a star whose photometry is perfectly constant over time. Its reference flux is determined on a reference image with a given circular aperture radius. In another image where the PSF FWHM is larger, the star's flux will remain the same, but will be spread-out because of degraded seeing. If we use the circular aperture convolved with the kernel $K$ as the new aperture, we can understand why the outer pixel values (that should contribute to the true photometry) will be artificially decreased, whereas the background noise, in areas where it becomes dominant compared to the star signal, will be further attenuated.

Thus the image subtraction algorithms intrinsically and globally produces more accurate (cleaner) lightcurves than aperture photometry, but has a tendency to underestimate the photometry (and the noise level) in the case where the above-described aperture convolution procedure is used.

It should be pointed out that the photometric bias is more important for fainter stars due to the combination of two effects: firstly, as mentioned above, the associated reference image aperture radius will be smaller than for a brighter star, and secondly because the kernel size is approximately constant over the entire image. Since the average kernel size directly depends on the seeing, the photometric underestimation bias will be even stronger for fainter stars during bad seeing episodes, and can reach up to about 30\% in our data.

In order to compensate for this photometric bias, we need to introduce a correction factor $\alpha$ such that,

\begin{equation}
F = F_\mathrm{ref} + \alpha\,\frac{(A \otimes K)\,[I - (R \otimes K)]} {\parallel K \parallel^2}\,\,.
\end{equation}

\subsection{Photometric correction factor $\alpha$}
\begin{figure}
    \centering
    \begin{minipage}[c]{\columnwidth}
        \centering\includegraphics[width=\columnwidth]{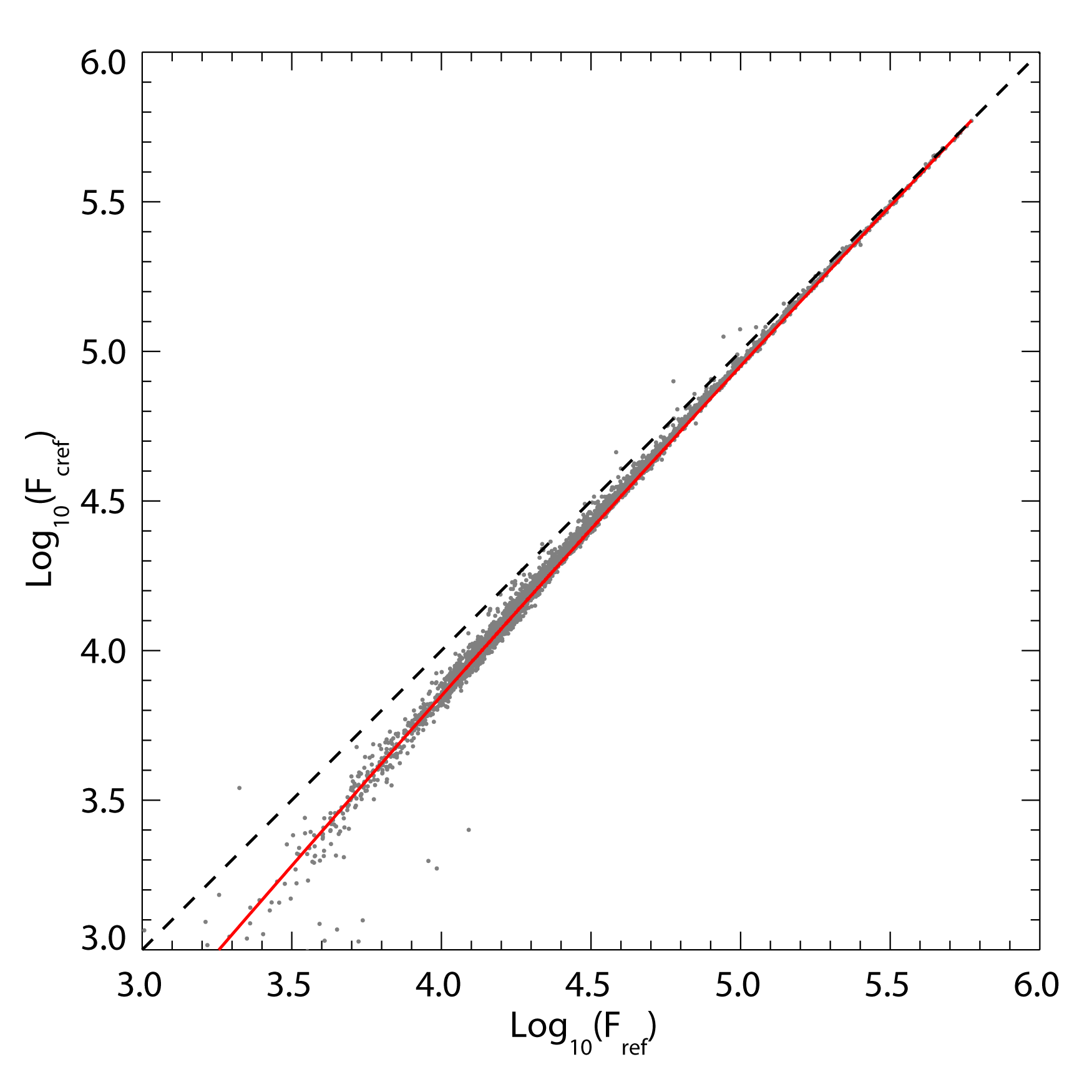}
    \end{minipage}
    \caption{\label{fig:photometric.calibration}Plot of the reference image fluxes computed with a
circular aperture ($F_\mathrm{ref}$) versus the same reference image computed with the
kernel-convolved aperture $A \otimes K$ ($F_\mathrm{cref}$). The fitted curve is shown in red, as
well as the ideal linear trend (dashed line).}
\end{figure}

To assess the $\alpha$ factor, we simply plot the reference fluxes $F_\mathrm{ref}$ of each star against the reference PSFs, convolved by the modified aperture, that is $F_\mathrm{cref}=(A \otimes K)\,(R \otimes K)\, / \parallel K \parallel^2$. This shows how the reference fluxes (in the reference image) compare to the fluxes when we use the convolved aperture $A \otimes K$. If the field contains enough stars, a fitting curve can be adjusted in order to give the photometric compensation $\alpha$ for any measured photometric residual (Fig.\,\ref{fig:photometric.calibration}). The photometric correction factor $\alpha$ needs to be computed for every image since the optimal kernels are changing with each image. \textcolor{referee}{The uncertainty on $\alpha$ was evaluated by scrambling all the points, randomly switching them (or not) with its immediate neighbor, then performing the fit and repeating the procedure 1000 times. This test shows that $\alpha$ is stable at the level of about 1\,\% up to magnitude $R=16$, and raising to about 5\,\% at $R=18$. An additional test showing the robustness of the method is that the correction factor $\alpha$ is very well correlated with the PSF FWHM (thus consistent with the explanation given above). Although we cannot verify the exactness of the $\alpha$ factor for each of the stars, it appears to be an efficient first order photometric correction. Most importantly it does not introduce spurious noise structures in the lightcurves as it is the case of WASP-19 (see below).}

\subsection{The case of the WASP-19 field}
In the case of WASP-19, the transit photometry was underestimated by approximately 10\%. We also found that the described photometric correction procedure can be the source of additional noise (specifically for that field) if the correction factor is computed and applied to each individual image. Indeed, in that case, the correction is not simply a scaling factor, but seems to generate more high frequency fluctuations (typically of about an hour length). The reason is that for WASP-19, we applied the OIS on a small crop ($600\times600$ pixels) rather than on the full image to increase the OIS precision \textcolor{referee}{(recalling that the kernel is allowed to vary according to second order variations)}. The WASP-19 field is not as crowded as other of our fields where the correction relies on much larger star counts. We therefore computed the photometric compensation factor for each image individually, and used a temporally smoothed version of it to compensate for the original OIS data. In that way, the original signal to noise is preserved, while correctly compensating for the photometric bias, mostly visible when deriving the primary transit parameters. Note that without compensating for the 10\% photometric bias, the obtained primary transit parameters are way beyond the ones obtained with the compensated data (including error bars). The compensated data produce parameters that are totally consistent with values derived in previous papers (e.g. \citet{Anderson2010}). Especially in \citet{Hellier2011} a WASP-19b transit lightcurve obtained with the NTT in the Gunn \textit{r}-band is shown (see their Fig.\,1--bottom), that was recorded just three months prior to our own observations, and is totally consistent with our own results.

\subsection{Comparison between aperture photometry and OIS\label{Appendix.Compare.OIS.AP}}
\begin{figure}
    \centering
    \begin{minipage}[c]{\columnwidth}
        \centering\includegraphics[width=\columnwidth]{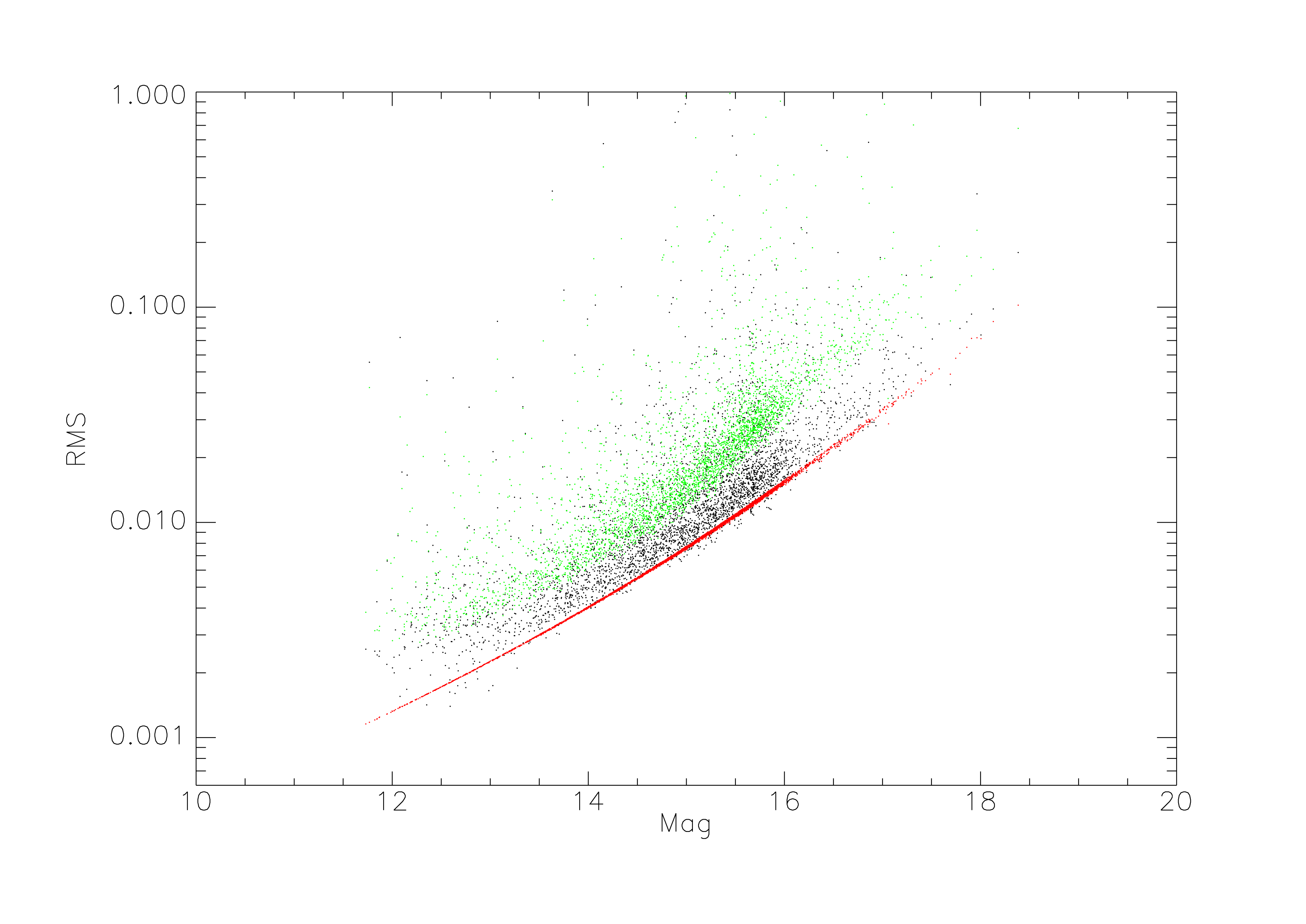}
    \end{minipage}
    \caption{\label{fig:compare.PO.OIS}Lightcurves RMS in the WASP-19 field for the aperture photometry (green dots) and for the OIS (black dots) taking into account the photometric correction described in Appendix\,\ref{Appendix.A}. The red dots correspond to the fundamental noise estimate (including photon, background and read noises).}
\end{figure}

Figure\,\ref{fig:compare.PO.OIS} is a comparison between the residuals obtained for both the aperture photometry and the OIS algorithms in the whole WASP-19 field. The lightcurve calibration procedure is identical for both algorithms (only the raw lightcurves are different).

\section{Day-to-day transit lightcurves\label{Appendix.LC}}
\begin{table*}
\caption{\label{tab:tbright}Basic data and primary transit parameters for each of the observing night.}
\begin{center}
\begin{spacing}{1.5}
\begin{tabular}{llcccccc}
\hline\hline
Night   & $\sharp$ data & $t_\mathrm{mid}$ & $t_{1,4}$ & $R_\mathrm{p}/R_{\bigstar}$ & \textbf{P}ink$_\mathrm{noise}$ & \textbf{W}hite$_\mathrm{noise}$ & \textbf{R}ed$_\mathrm{noise}$\\
        &               & (HJD) & (min) &  & (ppm) & (ppm) & (ppm)\\
\hline
2        & 257    & 2455318.05715  & 106.25        & 0.150 & 722 & 429 & 580\\
3        & 275    & --             & --            & --    & 434 & 392 & 186\\
4        & 273    & 2455320.42864  &  76.44        & 0.135 & 661 & 439 & 494\\
5        & 284    & 2455321.21524  &  99.19        & 0.145 & 594 & 342 & 486\\
6        & 293    & 2455322.00248  &  95.50        & 0.139 & 484 & 429 & 224\\
7        &  24    & --             & --            & --    & --  & --  & -- \\
8        & 217    & --             & --            & --    & 769 & 426 & 640\\
9        & 286    & 2455325.15851  & 100.06        & 0.135 & 776 & 374 & 680\\
10       & 230    & 2455325.94792  &  94.95        & 0.134 & 683 & 571 & 375\\
11       & 286    & --             & --            & --    & 415 & 342 & 236\\
12       & 341    & 2455328.31349  &  96.78        & 0.139 & 734 & 326 & 657\\
13       & 311    & 2455329.10318  &  95.34        & 0.140 & 962 & 409 & 871\\
14       & 260    & --             & --            & --    & 730 & 370 & 629\\
15       & 351    & 2455331.46889  & 108.73        & 0.155 & 728 & 349 & 639\\
16       & 295    & 2455332.25741  &  97.74        & 0.140 & 646 & 332 & 554\\
17       & 345    & 2455333.04710  & 106.19        & 0.147 & 595 & 340 & 489\\
18       & 284    & 2455333.83724  &  99.80        & 0.144 & 708 & 412 & 576\\
19       & 360    & 2455318.05715  & 106.25        & 0.150 & 441 & 344 & 275\\
20       & 127    & --             & --            & --    & 741 & 363 & 646\\
21       & 367    & 2455336.99102  &  95.57        & 0.139 & 591 & 372 & 459\\
22       & 359    & --             & --            & --    & 499 & 338 & 367\\
23       & 377    & 2455339.35751  &  95.94        & 0.139 & 955 & 389 & 872\\
24       & 338    & 2455340.14610  &  95.77        & 0.133 & 647 & 435 & 480\\
\hline
\end{tabular}
\end{spacing}
\end{center}
\end{table*}

\begin{figure*}
    \centering
    \begin{minipage}[c]{\textwidth}
        \centering\includegraphics[width=\textwidth]{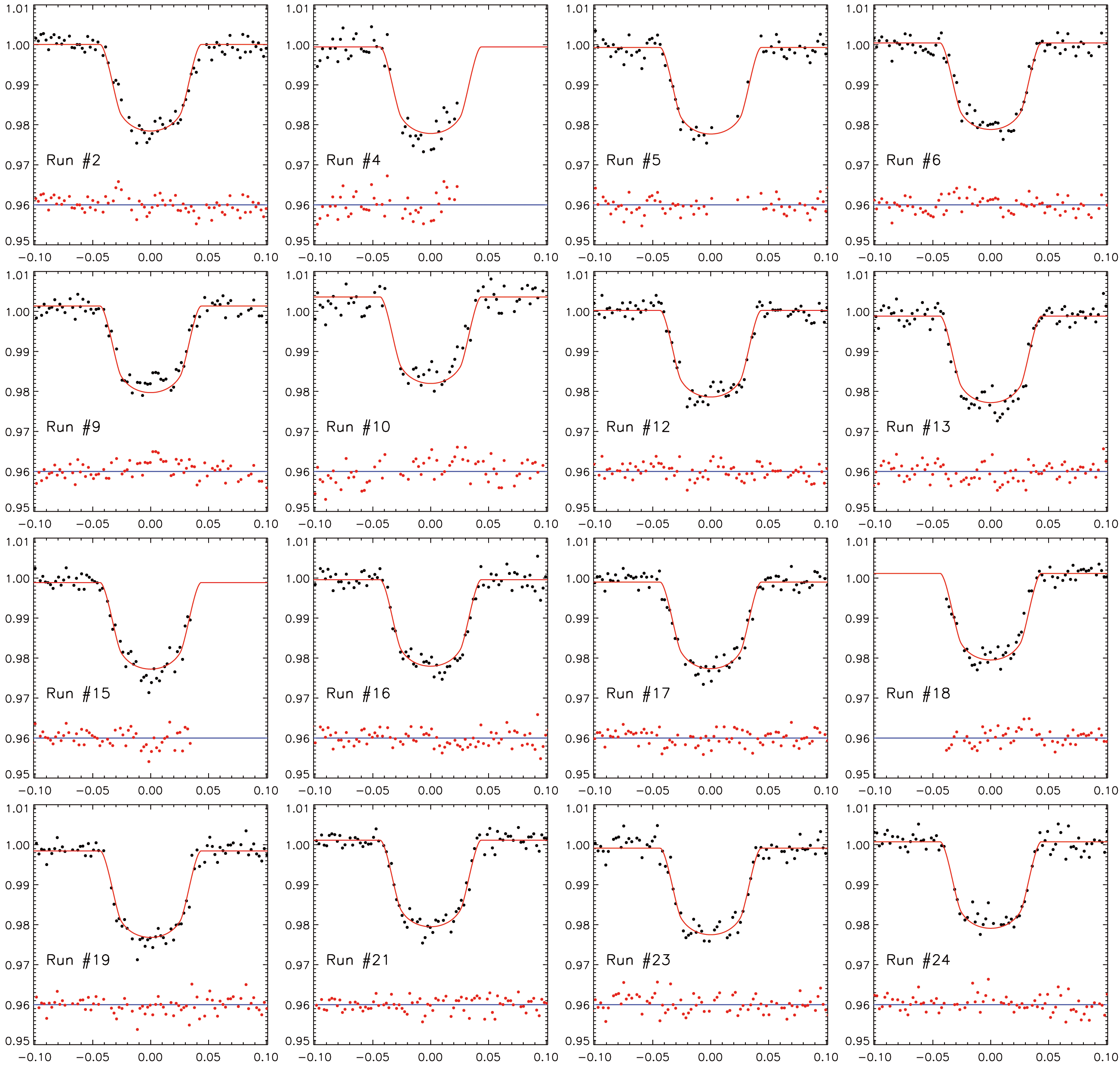}
    \end{minipage}
    \caption{\label{fig:day.to.day.transits}Day-to-day lightcurves of the primary transits events, along with the best-fit model (from parameters of Table\,\ref{Table:WASP19_TransitParam}) overplotted in red, and the residual points, shifted in ordinate for clarity.}
\end{figure*}

\end{appendix}

\begin{acknowledgements}
The ASTEP project has been funded by the French {\em Agence Nationale de la Recherche} (ANR), {\em Institut des Sciences de l'Univers} (INSU), {\em Programme National de Plan\'etologie}, \textit{Institut Paul-\'{E}mile Victor} (IPEV), and the {\em Plan Pluri-Formation} OPERA between {\em Observatoire de la C\^ote d'Azur} and {\em Universit\'e de Nice Sophia Antipolis} (UNS). Ivan Gon\c{c}alves is funded by INSU/CNRS. T.G. acknowledges support from a Fulbright fellowship while
at UCSC.
\end{acknowledgements}


\begin{thebibliography}{}
\bibitem[{{Alard} \& {Lupton}(1998)}]{Alard1998} {Alard}, C., \& {Lupton}, R.H.\ 1998 ApJ 503, pp.\,325-331
\bibitem[{{Alonso} {et~al.}(2009a){Alonso}, {Guillot}, {Mazeh}, {Aigrain}, {Alapini}, {Barge}, {Hatzes}, {Pont}}]{Alonso2009a} {Alonso}, R., {Guillot}, T., {Mazeh}, T. {et~al.}\ 2009, \aap, 501, L23
\bibitem[{Alonso} {et~al.}(2009b)]{Alonso2009b}  {Alonso}, R., {Alapini}, A., {Aigrain}, S. {et~al.}\ 2009, \aap, 506, 353
\bibitem[{{Andersen} {et~al.}(1995){Andersen}, {Freyhammer}, {Storm}}]{Andersen1995} {Andersen}, M.~I., {Freyhammer}, L., \& {Storm}, J.\ 1995, European Southern Observatory Conference and Workshop Proceedings, 53, 87
\bibitem[{{Anderson} {et~al.}(2010)}]{Anderson2010} {Anderson}, D.~R., {Gillon}, M., {Maxted}, P.~F.~L. {et~al.}\ 2010, \aap, 513, L3
\bibitem[{{Anderson} {et~al.}(2011)}]{Anderson2011} Anderson, D.~R., Smith, A.~M.~S., Madhusudhan, N. {et~al.}\ 2011, arXiv:1112.5145
\bibitem[{{Barker} \& {Ogilvie}(2009)}]{Barker2009} Barker, A.~J., \& Ogilvie, G.~I.\ 2009, \mnras, 395, 2268
\bibitem[{{Barker} \& {Ogilvie}(2010)}]{Barker2010} Barker, A.~J., \& Ogilvie, G.~I.\ 2010, \mnras, 404, 1849
\bibitem[{{Borucki} {et~al.}(2009)}]{Borucki2009} Borucki, W.~J., Koch, D., Jenkins, J., et al.\ 2009, Science, 325, 709
\bibitem[{{Bouvier} {et~al.}(1997)}]{Bouvier1997} Bouvier, J., Forestini, M., \& Allain, S.\ 1997, \aap, 326, 1023
\bibitem[{{Burton} {et~al.}(2012)}]{Burton2012} Burton, J.~R., Watson, C.~A., Littlefair, S.~P., et~al.\ 2012, \apjs, 201, 36
\bibitem[{{Carone} \& P\"{a}tzold (2007)}]{Carone2007} Carone, L., P\"{a}tzold, M.\ 2007, \planss, 55, 643
\bibitem[{{Claret} \& {Bloemen}(2011)}]{Claret2011} Claret, A., \& Bloemen, S.\ 2011, VizieR Online Data Catalog, 352, 99075
\bibitem[{{Christiansen} {et~al.}(2010)}]{Christiansen2010} Christiansen, J.~L., Ballard, S., Charbonneau, D., et~al.\ 2010, \apj, 710, 97
\bibitem[{{Cowan} \& {Agol}(2011)}]{Cowan2011} Cowan, N.~B., \& Agol, E.\ 2011, \apj, 729, 54
\bibitem[{{Crouzet} {et~al.}(2010)}]{Crouzet2010} Crouzet, N., Guillot, T., Agabi, A. et~al.\ 2010, \aap, 511, A36
\bibitem[{{Daban} {et~al.}(2010)}]{Daban2010} Daban, J.-B., et al. 2010, \procspie, 7733
\bibitem[{{Deming} {et~al.}(2005)}]{Deming2005} Deming, D., Seager, S., Richardson, L.~J., \& Harrington, J.\ 2005, \nat, 434, 740
\bibitem[{{Fortney} {et~al.}(2005){Fortney}, {Marley}, {Lodders}, {Saumon}, \& {Freedman}}]{Fortney05}{Fortney}, J.~J., {Marley}, M.~S., {Lodders}, K., {Saumon}, D., \& {Freedman}, R. 2005, \apjl, 627, L69
\bibitem[{{Fortney} {et~al.}(2008){Fortney}, {Lodders}, {Marley}, \& {Freedman}}]{Fortney08}{Fortney}, J.~J., {Lodders}, K., {Marley}, M.~S., \& {Freedman}, R.~S. 2008, \apj, 678, 1419
\bibitem[{{Gibson} {et~al.}(2010)}]{Gibson2010} Gibson, N.~P., Aigrain, S., Pollacco, D.~L. et~al.\ 2010, \mnras, 404, L114
\bibitem[{{Gillon} {et~al.}(2006)}]{Gillon2006} Gillon, M., Pont, F., Moutou, C. et~al.\ 2006, \aap, 459, 249-255
\bibitem[{{Gillon} {et~al.}(2007)}]{Gillon2007} Gillon, M., Pont, F., Moutou, C. et~al.\ 2007, \aap, 466, 743
\bibitem[{{Guillot} \& {Havel}(2011)}]{Guillot2011} Guillot, T., \& Havel, M.\ 2011, \aap, 527, A20
\bibitem[{{Harrington} {et~al.}(2006)}]{Harrington2006} Harrington, J., Hansen, B.~M., Luszcz, S.~H., et~al.\ 2006, Science, 314, 623
\bibitem[{{Hauschildt} {et~al.}(1998)}]{Hauschildt1998} Hauschildt, P.~H., Aufdenberg, J., Starrfield, S., \& Baron, E.\ 1998, Wild Stars in the Old West, 137, 96
\bibitem[{{Hebb} {et~al.}(2010)}]{Hebb2010} Hebb, L., Collier-Cameron, A., Triaud, A.~H.~M.~J. et~al.\ 2010, \apj, 708, 224
\bibitem[{{Hellier} {et~al.}(2011)}]{Hellier2011} Hellier, C., Anderson, D.~R., Collier-Cameron, A. et~al.\ 2011, \apjl, 730, L31
\bibitem[{Husnoo} {et~al.}(2012)]{Husnoo2012} Husnoo, N., Pont, F., Mazeh, T., Fabrycky, D., H{\'e}brard, G., Bouchy, F., \& Shporer, A.\ 2012, \mnras, 422, 3151
\bibitem[{{Kipping} \& {Bakos}(2011)}]{KippingBakos2011} Kipping, D., \& Bakos, G.\ 2011, \apj, 733, 36
\bibitem[{{Knutson} {et~al.}(2007)}]{Knutson2007} Knutson, H.~A., Charbonneau, D., Allen, L.~E., et al.\ 2007, \nat, 447, 183
\bibitem[{{Mandel} {et~al.}(2002)}]{Mandel2002} Mandel, K., \& Agol, E., 2002, \apj, 580, 171
\bibitem[{{Mazeh} {et~al.}(2009)}]{Mazeh2009} Mazeh, T., Guterman, P., Aigrain, S., et al.\ 2009, \aap, 506, 431
\bibitem[{{Meibom} \& {Mathieu}(2005)}]{Meibom2005} Meibom, S., \& Mathieu, R.~D.\ 2005, \apj, 620, 970
\bibitem[{{Miller} {et~al.}(2008)}]{Miller2008} Miller, J.~P., Pennypacker, C.~R., \& White, G.~L.\ 2008, \pasp, 120, 449
\bibitem[{{Montalto} {et al.}(2007)}]{Montalto2007}{Montalto}, M., {Piotto}, G., {Desidera}, S., et~al.\ 2007, \aap, 470, 1137
\bibitem[{{Morel} \& {Lebreton}(2008)}]{Morel2008} Morel, P., \& Lebreton, Y.\ 2008, \apss, 316, 61
\bibitem[{{Ogilvie} \& {Lin}(2007)}]{Ogilvie2007} Ogilvie, G.~I., \& Lin, D.~N.~C.\ 2007, \apj, 661, 1180
\bibitem[{{Pont} {et~al.}(2006)}]{Pont2006} Pont, F., Zucker, S., \& Queloz, D.\ 2006, \mnras, 373, 231
\bibitem[{{Rowe} {et~al.}(2008)}]{Rowe2008} Rowe, J.~F., Matthews, J.~M., Seager, S., et al.\ 2008, \apj, 689, 1345
\bibitem[{{Saumon} {et~al.}(1996)}]{Saumon1996} Saumon, D., Hubbard, W.~B., Burrows, A., Guillot, T., Lunine, J.~I., \& Chabrier, G.\ 1996, \apj, 460, 993
\bibitem[{{Showman} \& {Guillot}(2002)}]{ShowmanGuillot2002} Showman, A.~P., \& Guillot, T.\ 2002, \aap, 385, 166
\bibitem[{{Snellen} {et~al.}(2009)}]{Snellen2009} Snellen, I.~A.~G., de Mooij, E.~J.~W., \& Albrecht, S.\ 2009, \nat, 459, 543
\bibitem[{{Snellen} {et~al.}(2010)}]{Snellen2010} Snellen, I.~A.~G., de Mooij, E.~J.~W., \& Burrows, A.\ 2010, \aap, 513, A76
\bibitem[{{Tamuz}, {Mazeh}, \& {Zucker}(2005)}]{Tamuz2005} Tamuz, O., Mazeh, T., \& Zucker, S.\ 2005, \mnras, 356, 1466
\bibitem[{{Torres}, {et~al.}(2008)}]{Torres2008} Torres, G., Winn, J.~N., \& Holman, M.~J.\ 2008, \apj, 677, 1324

\end{thebibliography}
\end{document}